\documentclass[apj]{emulateapj}
\usepackage{apjfonts}
\usepackage{epsf}
\def\beq{\begin{equation}} 
\def\enq{\end{equation}}   
\newcommand{\bdv}[1]{\mbox{\boldmath$#1$}}
\newcommand{\Msun}{\mathrm{M}_{\odot}}

\begin{document}

\slugcomment{Submitted to ApJ}
\shortauthors{PRIETO \& GNEDIN}
\shorttitle{Dynamical Evolution of Globular Clusters}

\title{Dynamical Evolution of Globular Clusters in Hierarchical Cosmology}

\author{Jos\'e L. Prieto and Oleg Y. Gnedin}

\affil{The Ohio State University,
       Department of Astronomy,
       140 W 18th Ave., Columbus, OH 43210 \\
       \mbox{\tt prieto@astronomy.ohio-state.edu,
                 ognedin@astronomy.ohio-state.edu}}

\begin{abstract}
We test the hypothesis that metal-poor globular clusters form within
disk galaxies at redshifts $z>3$.  Numerical simulations demonstrate
that giant gas clouds, which are cold and dense enough to produce
massive star clusters, assemble naturally in hierarchical models of
galaxy formation at high redshift.  Do model clusters evolve into
observed globular clusters or are they disrupted before present as a
result of the dynamical evolution?  To address this question, we
calculate the orbits of model clusters in the time-variable
gravitational potential of a Milky Way-sized galaxy, using the outputs
of a cosmological $N$-body simulation.  We find that at present the
orbits are isotropic in the inner 50 kpc of the Galaxy and
preferentially radial at larger distances.  All clusters located
outside 10 kpc from the center formed in satellite galaxies, some of
which are now tidally disrupted and some of which survive as dwarf
galaxies.  The spatial distribution of model clusters is spheroidal
and the fit to the density profile has a power-law slope $\gamma
\approx 2.7$, somewhat shallower than but consistent with observations
of metal-poor clusters in the Galaxy.  The combination of two-body
relaxation, tidal shocks, and stellar evolution drives the evolution
of the cluster mass function from an initial power law to a peaked
distribution, in agreement with observations.  However, not all
initial conditions and not all evolution scenarios are consistent with
the observed mass function of the Galactic globular clusters.  The
successful models require the average cluster density, $M/R_h^3$, to
be constant initially for clusters of all mass and to remain constant
with time.  Synchronous formation of all clusters at a single epoch
($z=4$) and continuous formation over a span of 1.6 Gyr (between $z=9$
and $z=3$) are both consistent with the data.  For both formation
scenarios, we provide online catalogs of the main physical properties
of model clusters.
\end{abstract}


\section{Introduction}
  \label{sec:intro}

Observations in the past decade with the Hubble Space Telescope and
major ground-based telescopes have revolutionized the field of star
cluster research.  The classical division into old, massive globular
clusters and young, diffuse open clusters has been challenged by the
discoveries of young massive star clusters in nearby merging,
interacting, and even normal star forming galaxies
\citep[e.g.,][]{holtzman_etal92, whitmore_etal93, whitmore_schweizer95,
oconnell_etal95, zepf_etal99, larsen02}.  These young clusters have the
range of masses and sizes similar to Galactic globulars and are believed
to evolve eventually into globular clusters
\citep[e.g.,][]{ho_filippenko96, mengel_etal02, degrijs_etal04}.
However, the mass functions of old and young clusters are very
different.  While the mass distribution of globular clusters has a
well-defined peak at $\approx 2\times 10^5\, \Msun$, the mass function
of young clusters is consistent with a single power law between $10^4$
and $10^7\, \Msun$ \citep{zhang_fall99, degrijs_etal03, anders_etal04}.
{\it Did old globular clusters form by a similar physical process to the
young clusters but gradually evolve into the observed distribution?}

This question is vital to understanding early galaxy formation and is
still a matter of debate \citep{fall_zhang01,vesperini_etal03}.
Globular clusters contain relic information about star formation
processes in galaxies at high redshift and have been often used to
constrain cosmological theory.  Uncovering the formation of primeval
galaxies is a major driving force behind the development of future
major ground and space observatories.  In contrast to
the formidable difficulties of detecting the birthplaces of globular
clusters at high redshift, local and interacting galaxies (M33, M51,
M82, the Antennae) offer plenty of detail of the formation of young
star clusters in giant molecular clouds \citep{wilson_etal03,
engargiola_etal03, keto_etal05}.  Using ultrahigh-resolution
cosmological simulation, \citet{kravtsov_gnedin05} identified similar
molecular clouds within their simulated high-redshift galaxies and
proposed that they harbor massive star clusters.  In this paper, we
investigate whether the dynamically-evolved distribution of these
model clusters can be reconciled with the current properties of
globular clusters.

\citet{kravtsov_gnedin05} find that giant clouds assemble during
gas-rich mergers of progenitor galaxies, when the available gas forms a
thin, cold, self-gravitating disk.  The disk develops strong spiral
arms, which further fragment into separate molecular clouds located
along the arms as beads on a string (see their Fig. 1).  As a result,
massive clusters form in relatively massive galaxies, with the total
mass $M_{\rm halo} > 10^{9}\ \Msun$, beginning at redshift $z \approx
10$.  The mass and density of the molecular clouds increase with cosmic
time, but the rate of galaxy mergers declines steadily.  Therefore, the
cluster formation efficiency peaks at a certain extended epoch, around
$z \approx 4$, when the Universe is only 1.5 Gyr old.  The mass function
of model clusters is consistent with a power law
\begin{equation}
  dN/dM = N_0 \, M^{-\alpha},
  \label{eq:dndm}
\end{equation}
where $\alpha = 2.0 \pm 0.1$, similar to the observations of nearby
young star clusters.  According to the prescription for metal enrichment
by supernovae in the \citet{kravtsov_gnedin05} simulation, model
clusters have iron abundances $[Fe/H] \le -1$ at $z > 3$.  They
correspond to the metal-poor (dominant) sub-population of the Galactic
globular clusters, which we will use for the comparison.

We adopt this model to set up the initial positions, velocities, and
masses for our globular clusters.  We then calculate cluster orbits
using the results of a separate collisionless $N$-body simulation
described in \S~\ref{sec:orbits_cl}.  This is necessary because the
original gasdynamics simulation was stopped at $z \approx 3$, due to
limited computational resources.  By using the $N$-body simulation of
a similar galactic system, but complete to $z=0$, we are able to
follow the full dynamical evolution of globular clusters until the
present epoch.  We use the evolving properties of all progenitor
halos, from the outputs with a time resolution of $\sim 10^8$ yr, to
derive the gravitational potential in the whole computational volume
at all epochs.  We convert a fraction of the dark matter mass into the
analytical flattened disks, in order to model the effect of baryon
cooling and star formation on the galactic potential.  We calculate
the orbits of globular clusters in this potential from the time when
their host galaxies accrete onto the main (most massive) galaxy.
Using these orbits, we calculate the dynamical evolution of model
clusters, including the effects of stellar mass loss, two-body
relaxation, tidal truncation, and tidal shocks.  Since the efficiency
of each of these processes depends on both cluster mass, $M$, and
half-mass radius, $R_h$, we consider several evolutionary dependencies
$R_h(M)$.  We also consider two possible formation scenarios, one with
all clusters forming in a short interval of time around redshift
$z=4$, and the other with a continuous formation of clusters between
$z=9$ and $z=3$.

Note that we study only the collisionless evolution of star clusters,
using the analytical formalism developed in previous studies and
described in \S~\ref{sec:destruction}.  We do not include the recently
suggested ``infant mortality'' effect, which may dissolve a large
number of unbound clusters within $\sim 100$ Myr of their formation
independent of cluster mass \citep[e.g.,][]{fall_etal05,
mengel_etal05}.  Such universal reduction in numbers is therefore
absorbed in our normalization of the initial GCMF.  The models
calculated in this paper apply only to the initially bound clusters.

\section{Orbits of Globular Clusters in a Hierarchically-Forming Galaxy}
  \label{sec:orbits_cl}

\subsection{Cosmological Simulation}
  \label{sec:simul} 

We set the initial conditions and calculate the orbits of model clusters
using a cosmological $N$-body simulation performed with the Adaptive
Refinement Tree code \citep[ART,][]{kravtsov_etal97, kravtsov99}.  This
simulation follows the hierarchical formation of a Milky Way-sized halo
in a comoving box of $25h^{-1}$~Mpc in the concordance $\Lambda$CDM
cosmology: $\Omega_{\rm m}=0.3$, $\Omega_{\Lambda}=0.7$,
$\sigma_{8}=0.9$, $H_{0}=70$ km s$^{-1}$ Mpc$^{-1}$.  The simulation
begins with a uniform $256^3$ grid covering the entire computational
volume.  High mass resolution is achieved around collapsing structures
by recursive refinement of this regions using an adaptive refinement
algorithm.  The grid cells are refined if the particle mass contained
within them exceeds a certain threshold value.  Thus, this adaptive grid
follows the collapsing galaxy in a quasi-lagrangian fashion.  With
maximum 10 levels of refinement, the peak formal spatial resolution is
0.15 comoving kpc.  This simulation is described in more detail in
\citet{kravtsov_etal04}.

Halos and subhalos are identified using a variant of the Bound Density
Maxima algorithm \citep{klypin_etal99}.  The halo finder algorithm
produces halo catalogs, which contain positions, velocities, and main
properties of the dark matter halos between $z=9$ and $z=0$ at 96
output times, with a typical separation between outputs $\approx 10^{8}$
yr.  The main properties of the halos in the catalogs are: the virial
radius ($r_{\rm vir}$), truncation radius ($r_{\rm t}$), halo radius
($r_{\rm halo} = \min [r_{\rm t}, r_{\rm vir}]$), halo mass ($M_{\rm
halo}$), maximum circular velocity ($V_{\rm max}$) and the corresponding
radius at which the maximum velocity is reached ($r_{\rm max}$).  The
virial radius (and corresponding virial mass) is defined as the radius
within which the density is 180 times the mean density of the
Universe. The truncation radius is the radius at which the logarithmic
slope of the density profile becomes larger than -0.5, since we do not
expect the density profile of the halos to be flatter than this slope.
If the center of a halo does not lie within the boundary of a larger
system, its radius $r_{\rm halo}$ is defined equal to the virial radius.
On the other hand, if the halo is within another halo (and is therefore
a subhalo), its radius is the truncation radius.  The halo mass follows
the same definition: $M_{\rm halo} = \min [M(r_{\rm t}), M(r_{\rm
vir})]$.

In this paper, we consider the evolution of an isolated halo G$_1$,
which we call the {\em main halo}.  This halo contains over $10^6$
particles within its virial radius at $z=0$.  The virial mass at $z=0$
in our definition is $M_{\rm vir} = 2.37\times 10^{12}\, \Msun$, the
virial radius $r_{\rm vir} = 426$~kpc and the halo concentration $c=13$.
For the commonly used overdensity of 340, the virial mass and radius are
$M_{340} = 2.07\times 10^{12}\, \Msun$ and $r_{340} = 330$ kpc,
respectively.  The main halo has experienced the last major merger at $z
\approx 2$ and therefore could subsequently host a large spiral galaxy.

The main advantage of the catalogs of \citet{kravtsov_etal04} is that
they trace the progenitors of halos at previous epochs and therefore
allow us to reconstruct the full dynamical evolution of the model
galaxies.  This information is necessary to follow the evolution of
globular clusters within their host galaxies, before being accreted onto
the main halo.

\subsection{Galactic Potential}
  \label{sec:potential}

We construct realistic time-dependent galactic potentials using the halo
catalogs and supplementing each massive dark matter halo with a baryonic
disk.  The distribution of dark matter in a cosmological simulation can
be approximately described as a combination of isolated halos and
satellites within them, and therefore the gravitational potential can be
approximated by a sum of the potentials of the main halo and all
subhalos \citep{kravtsov_etal04}.  For each galaxy (isolated and
satellite) we assign a fraction of the mass, $M_{\rm disk}= f_{\rm d} \,
M_{\rm halo}$, to the baryonic disk and the remaining mass, $(1-f_{\rm
d}) M_{\rm halo}$, to the dark halo.  In this crude approximation, the
fraction $f_{\rm d}$ accounts for the potential of stars and cold gas in
the galaxy and absorbs all the complex details of star formation.  We do
not explicitly consider the galactic bulge because most cluster orbits
do not come too close to the bulge, and its potential can be described
by the inner part of the disk.  The bulge mass is therefore absorbed in
our definition of the disk mass.  The total potential, as a function of
space and time, is the sum of the halo and disk contributions of all the
galaxies:
\beq
  \Phi_{\rm}(\bdv{r},t) = \sum_{i} \Phi_{\rm halo} ^{i}(\bdv{r} -
    \bdv{r}_{i}) + \Phi_{\rm disk}^{i}(\bdv{r} - \bdv{r}_{i}),
  \label{eq:total_potential}
\enq 
where $\bdv{r}_{i}$ is the position of the center of mass of
halo $i$ at time $t$.

We assume that the dark matter potential of each galaxy is given by a
spherical NFW \citep{nfw97} profile:
\begin{equation}
  \Phi_{\rm halo}(r) = -{G M_{\rm halo} (1-f_d) \over r}
    {\ln{(1 + r/r_{\rm s})} \over \ln{(1+c)}-c/(1+c)}
  \label{eq:NFW_potential}
\end{equation}
where $c = r_{\rm halo}/r_{\rm s}$ is the halo concentration, and
$r_s$ is the scale radius approximated as $r_s\approx r_{\rm
max}/2.16$.  We use this latter relation because the profiles of some
of the smaller subhalos are not well resolved in the simulation and
the radius $r_{\rm max}$ is better constrained than $r_s$.  The
axisymmetric disk potential is modeled with a \citet{miyamoto_nagai75}
profile, in cylindrical coordinates ($R,Z$):
\beq
  \Phi_{\rm disk}(R,Z) = -\frac{G M_{\rm disk}}{\sqrt{R^2 + (a_{\rm d}+\sqrt{Z^2 + b_{\rm d}^2})^2}}
  \label{eq:MN_potential}
\enq 
where $M_{\rm disk}(t)$, $a_{\rm d}(t)$ and $b_{\rm d}(t)$ are the
mass, radial scale length, and vertical scale height of the disk,
respectively, which vary with time.

The masses and sizes of the halos are linearly interpolated between the
output epochs in the halo catalog.  For the parameters of the disk, we
apply the following simple model.  The disk mass increases linearly in
time from the initial epoch $z=9$ to the redshift of accretion $z_{\rm
acc}$.  This is defined as the epoch when the halo is accreted onto the
main halo, i.e., when its center of mass is within the virial radius of
the main halo.  We stop the growth of the disk when it is accreted and
keep its mass fixed at later times, regardless of the evolution of its
host halo.  For the main halo and other isolated halos, $z_{\rm acc}=0$
by definition.  Thus the initial disk mass is $f_{\rm d} \,M_{\rm
halo}(z=9)$ and the final mass is $f_{\rm d} \,M_{\rm halo}(z=z_{\rm
acc})$.  We set $f_{\rm d}=0.05$, consistent with the value inferred for
the Milky Way \citep{klypin_etal02}.

We use a similar recipe to obtain the scale length $a_{\rm d}(t)$ as a
function of time, by requiring $a_{\rm d} = 0.02 \, r_{\rm halo}$.  It
is interesting to note that $a_{\rm d}$ for the \citet{miyamoto_nagai75}
disk is equal within $3\%$ to the exponential scale length, $r_{\rm d}$,
defined as the radius at which the surface density decreases by a factor
$\Sigma(0)/\Sigma(r_{\rm d}) \equiv e$.  The ratio of the vertical scale
height to the radial scale length is set constant at all times, $b_{\rm
d}/a_{\rm d} = 0.2$, consistent with observed galaxy disks
\citep[e.g.,][]{binney_tremaine87}.  In our model, the disk of the main
galaxy has the mass $M_{\rm disk} = 1.2\times 10^{11}\, \Msun$ and scale
length $a_{\rm d} = 8.5$ kpc at $z=0$. Both of these parameters are
probably a factor of 2 larger than in the Milky Way, as is the total
halo mass.

We use {\em proper} (not comoving) coordinates and a reference frame
where the main halo is at the origin of the coordinate system at all
times.  With the potential given by
equation~(\ref{eq:total_potential}), the acceleration of a test
particle at ($\bdv{r}$,$t$) is calculated analytically:
\beq
  \bdv{a} = - \bdv{\nabla}_{r} \Phi_{\rm} + \Omega_{\Lambda} \, H_0^2 \, \bdv{r}.
  \label{eq:total_force}
\enq 
The second term in this equation is a constant acceleration per unit
length due to the dark energy, parametrized by the cosmological
constant \citep{lahav_etal91}.  It affects only clusters outside 200
kpc from the center.  We use cubic spline interpolation between the
outputs of the halo catalogs to obtain the positions of halos at
intermediate times.  The third order spline assures that the halo
velocities are smooth functions and the accelerations are continuously
defined at each catalog output.  However, this non-linear
interpolation of subhalo motions creates an additional acceleration of
the globular clusters within them, which needs to be included in
equation~(\ref{eq:total_force}).

\subsection{Initial Conditions for Globular Clusters}
  \label{sec:ic}

In the model of \citet{kravtsov_gnedin05}, globular clusters form in
dense cores of giant molecular clouds within the disks of high-redshift
galaxies.  We adopt this model to set the initial masses, positions, and
velocities of the clusters in each massive halo, $M_{\rm halo}> 10^9\,
\Msun$.  We use the initial globular cluster mass function (GCMF) given
by equation~(\ref{eq:dndm}), with the slope $\alpha=2$ and the
normalization such that the total mass in globular clusters in that
halo, $M_{GC}$, scales with the host halo mass as found in the
simulation of \citet{kravtsov_gnedin05}:
\begin{equation}
  M_{GC} = 3.2\times 10^6\, \Msun 
    \left( \frac{M_{\rm halo}}{10^{11}\, \Msun} \right)^{1.13}\times f.
  \label{eq:mass_GC}
\end{equation}
The additional scaling factor, $f \ge 1$, is needed for those models
where we assume that all clusters form at a single epoch rather than
over an extended period.  We choose this normalization factor such
that the number of surviving clusters at $z=0$ is approximately equal
to the number of observed metal poor clusters in the Galaxy.  For the
best model Sb-ii (see below) we have $f=12$; the continuous formation
model Cb-ii does not require a change in the normalization: $f=1$.  We
adopt the lower limit for the initial GCMF as $M_{\rm min} = 10^5\,
\Msun$ and the upper limit, $M_{\rm max}$, given by the fit of
\citet{kravtsov_gnedin05} scaled by the same factor $f$.
% \begin{equation}
%  M_{\rm max} = 2.9\times 10^6\, \Msun 
%    \left({M_{\rm halo} \over 10^{11}\, \Msun} \right)^{1.29},
% \end{equation}
The choice of the lower mass limit is motivated by our disruption
calculations, which show that no cluster of lower mass is expected to
survive the dynamical evolution over 10 Gyr (see
\S~\ref{sec:mass_function}).  The actual number of clusters in each host
galaxy is given by the discretization of the mass function.

After being accreted into the main halo, some host halos are tidally
disrupted.  There are a total of 77 halos that host globular clusters
within the virial radius of the main galaxy, of which 37 are disrupted
and 40 survive until $z=0$.  80\% of all host halos are accreted in the
redshift range $z_{\rm acc} \approx 0.2 - 1.9$.  The surviving
halos are accreted systematically later than the disrupted halos.  The
median redshift of accretion for all halos is 1.2, while for the
surviving subhalos it is 0.64.

We assume that the spatial number density of globular clusters follows
the distribution of baryon mass in their host galaxy disk: $dN(R)
\propto \Sigma_{\rm d}(R)\, R\, dR$, where $\Sigma_{\rm d}(R)$ is the
disk surface density.  We use this equation to obtain the initial radial
positions of the clusters, $R$.  We assign their azimuthal angles and
vertical positions randomly from the uniform distributions in the
intervals [0, $2\pi$] and [$-b_{\rm d}$, $+b_{\rm d}$], respectively.
There is no correlation between the initial cluster masses and
positions.  The clusters are set on circular orbits parallel to the
plane of the disk, with the initial velocities equal $V_{c} =
\sqrt{GM(R)/R}$, where $M(R)$ is the total halo plus disk mass within
radius $R$.  The initial velocities of both halos and globular clusters
include also the Hubble expansion velocity.

We consider two sets of cluster formation models:
\begin{enumerate}
\item {\it Single epoch of formation} (S). The clusters form at $z_f =
4$ in galaxies with $M_{\rm halo} > 10^9\, \Msun$.  The masses and
positions of clusters are generated using the masses and sizes of the
halos at $z_f$, but the velocities are obtained using the halo
properties at the epoch of accretion.  In the case of the main halo,
orbit integration starts at $z=4$.

\item {\it Continuous formation} (C).  Clusters form continuously in
the redshift range \mbox{$9 \geq z_f \geq 3$}, at each of the original
gasdynamics simulation outputs.  The average time interval between the
simulation outputs is $7 \times 10^7$ yr.  The span of cosmic time
between $z=9$ and $z=3$ is approximately 1.6 Gyr, which is consistent
the observed spread of ages of the metal-poor clusters in the Galaxy.
As in model (S), the masses and positions of clusters are obtained
using the halo properties at $z_f$ but the velocities are calculated
only at $z_{\rm acc}$.
\end{enumerate}

For each of these formation models, we consider several variations of
the initial relation between the cluster half-mass radius and its mass
and of the evolution of this relation with time.
% \footnote{In contrast to \citet{fall_zhang01}, we do not estimate $R_h$
% as a fixed fraction of the cluster tidal radius.  Tidal radius is
% determined by the instantaneous value of the tidal force along the
% cluster trajectory and therefore it varies significantly along the
% orbit.  This variation would spuriously change $R_h$ and affect the
% rates of cluster disruption.}
%
For the initial $R_{h} - M$ relation at the time of formation, we
consider two models:
\renewcommand{\labelenumi}{\alph{enumi}}
\begin{enumerate}
\item $R_{h}(0) = {\rm constant}$.  We use the median value for the
Galactic globular clusters, calculated using the online catalog of
\citet{harris96}: $R_{h} = 2.4$ pc.

\item $R_{h}(0) \propto M(0)^{1/3}$.  We use the constant average
half-mass density, $\rho_{h}(0) = 4\times 10^3\, \Msun \, {\rm
pc}^{-3}$, from the model of \citet{kravtsov_gnedin05} and set $R_{h}
= (3 M/4\pi \rho_{h})^{1/3}$.  The half-mass density does not depend
on the position of the cluster in its host galaxy, and therefore, by
construction does not reflect the local tidal field.  We make this
assumption for consistency with the formation model that we are
testing.
\end{enumerate}

For the time dependence of the $R_{h} - M$ relation, we consider three
variations: 
\renewcommand{\labelenumi}{\roman{enumi}}
\begin{enumerate}
\item $R_{h}(t) = {\rm constant}$, 

\item $R_{h}(t) \propto M(t)^{1/3}$, i.e. $\rho_{h}(t) = {\rm
constant}$,

\item $R_{h}(t) \propto M(t)$.
\end{enumerate}  

We construct 12 different model realizations by combining the single or
continuous formation of clusters (S, C) with the initial (a, b) and
evolutionary (i, ii, iii) size-mass relations.  In our notation, Sa-i is
the model with a single redshift of formation and constant half-mass
radius throughout the evolution; Cb-i has continuous formation in the
redshift range $9 \geq z \geq 3$, initially constant half-mass density,
and constant half-mass radius thereafter, etc.  These models are
summarized in Table~\ref{table:models1}.

For clusters that originate in galaxies that will have become satellites
of the main halo, we integrate the orbits beginning from the epoch of
accretion, $z_{\rm acc}$, rather than the epoch of formation, $z_f$.
The motivation for doing so is to avoid unnecessary computation of the
orbits of clusters inside their host galaxies at $z_f > z > z_{\rm
acc}$.  Before the host galaxies are accreted into the main halo, we
expect their globular clusters to remain on gravitationally-bound
equilibrium orbits.  For small hosts, the time step required to follow
these orbits is small, which leads to two problems.  One is the large
number of computations, the other is gradual accumulation of integration
errors.  The situation changes when the host galaxy becomes a satellite
-- globular clusters may experience significant tidal perturbations from
the main halo and may even escape their original host.  Therefore, we
assume that the clusters remain bound to their hosts at $z > z_{\rm
acc}$ and calculate their orbits only from $z = z_{\rm acc}$ until the
present.  We keep the cluster positions $R$ and $Z$ assigned at $z_f$
(equivalent to assuming initially circular orbits) but recalculate the
initial cluster velocities using the host mass at $z_{\rm acc}$.

For clusters in all models, we calculate the mass loss via two-body
relaxation and stellar evolution beginning at their assumed epoch of
formation, $z_f$.  We include tidal shocks only when we begin orbit
integration at $z_{\rm acc}$, because the mass loss rate due to tidal
shocks is very sensitive to the cluster orbits, as described in
\S~\ref{sec:destruction}.

%%%% Fig 1 %%%%%

\begin{figure*}[t]
\epsscale{1.0}
\vspace{-0.5cm}
\plotone{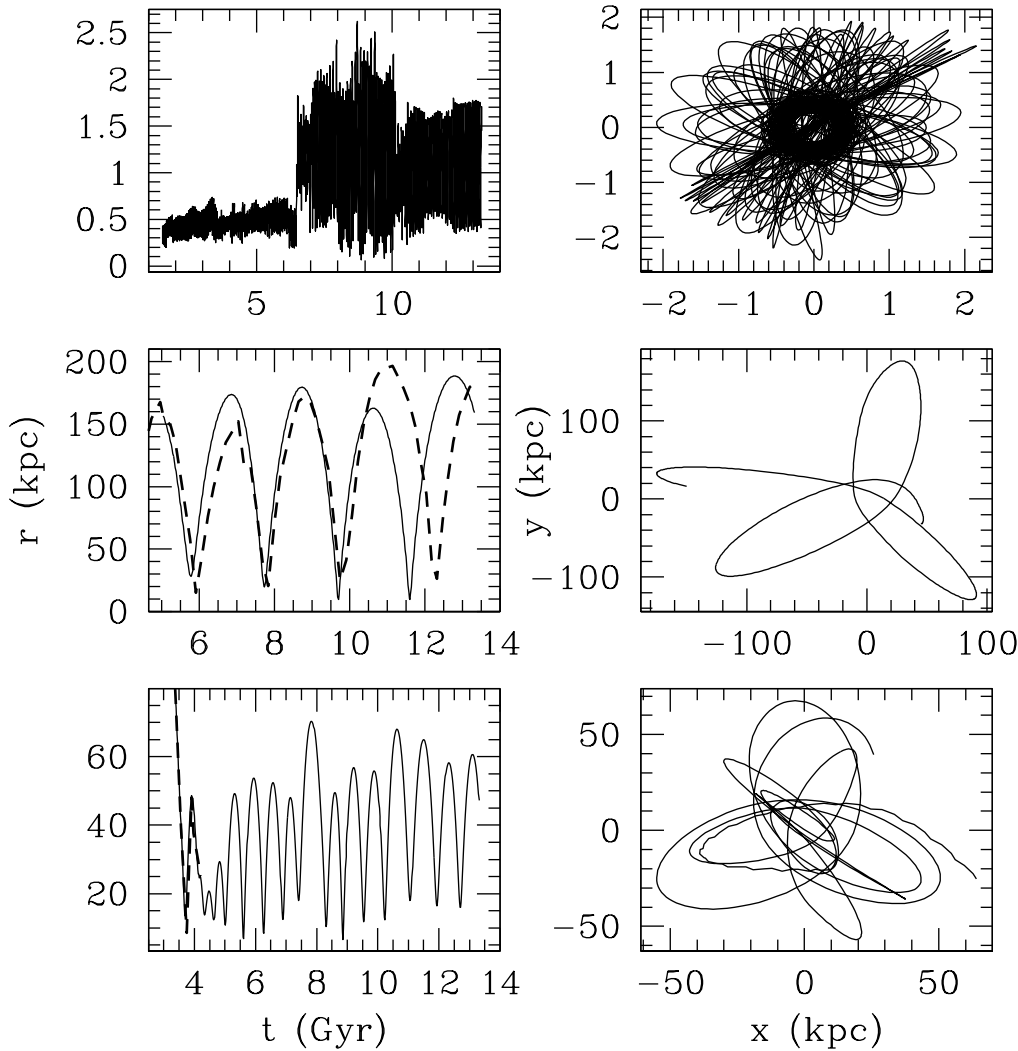}
\vspace{-0.2cm}
\caption{Three types of globular clusters orbits.  Left panels show
  the {\em proper} (not comoving) distance to the center of the main
  halo as a function of cosmic time, right panels show orbits in the
  plane of the main disk. {\em Top:} cluster formed in the main halo,
  on an initially circular orbit but was later scattered by accreted
  satellites.  {\em Middle:} cluster formed in a satellite halo, which
  survived as a distinct galaxy (halo is shown by a {\it dashed line}).
  {\em Bottom:} cluster formed in a satellite that was tidally
  disrupted at $t\simeq 4\ {\rm Gyr}$.}
  \label{fig:orbits_cl}
\end{figure*}

%%%%%%%%%%%%%%%%%%%

\subsection{Orbits of Host Halos and Globular Clusters}
  \label{sec:orbits}

The orbits of clusters in a time-dependent potential presented in
\S~\ref{sec:potential} are integrated using a second-order leapfrog
scheme with a fixed time step.  We have chosen this symplectic
integration scheme because it conserves energy to a precision of $\sim
10^{-10}$ in a static spherical potential.  A typical time step is
$10^4$ yr, which requires $\sim 10^3$ integration steps per cluster
orbit within a satellite host and $\sim 10^{4-5}$ steps per orbit
within the main halo.

Using only the outputs of an $N$-body simulation, we cannot reproduce
exactly the orbits that test objects would have had in that
simulation.  First, finite integration errors accumulate along the
orbits and diverge exponentially in time, such that two objects
initially close to each other may have quite different final
positions.  Second, we calculate the gravitational potential by
approximating the complex distribution of dark matter as a sum of
spherical halos.  This approximation is not entirely unreasonable and
reproduces the tidal field fairly accurately
\citep[see][]{kravtsov_etal04}, but of course it is not exact.  We
would like to test the orbit integration scheme on objects for which
we know their final positions in the simulation.  One class of such
objects are the satellite halos that host globular clusters.

We calculate the orbits of the host halos from $z_{\rm acc}$ to the
present and compare their predicted final radial distribution with the
actual distribution in the $N$-body halo catalogs.  In addition to the
gravitational potential given by equation~(\ref{eq:total_potential}),
we include the effect of dynamical friction, a drag force exerted by
the main halo on the orbiting satellites, using Chandrasekhar's
formula \citep{chandra43}:
\begin{equation}
  \bdv{a}_{\rm df} = - 4\pi G^2 M_{\rm halo} \, \rho \, \ln{\Lambda}
    \frac{\bdv{v}}{v^3} \left[ {\rm erf}(X) -
    \frac{2X}{\sqrt{\pi}}e^{-X^2}\right],
  \label{eq:DF}
\end{equation}
where $M_{\rm halo}$ is the mass of the satellite halo, $\rho(r)$ is
the local density of the main halo, $\bdv{v}$ is the velocity of the
satellite relative to the main halo, and $X \equiv v/[2\sigma(r)]$,
where $\sigma(r)$ is the one-dimensional velocity dispersion of
particles in the main halo, which was calculated assuming an isotropic
dispersion tensor using the approximate formula of
\citet{zentner_bullock03}.  Assuming a constant value for the Coulomb
logarithm, we find that with $\ln{\Lambda} = 5$
% 2.4 obtained by \citet{taylor_babul01} by comparing orbits of satellites
% from $N$-body simulations with their semianalytical models
the calculated radial distribution of satellite halos is similar to
that in the halo catalogs.  We somewhat overpredict the number of
halos at $r < 50$~kpc and underpredict this number at larger
distances, with the maximum deviation of $\sim 50\%$ at 150 kpc.  At
even larger distances the discrepancy decreases and we match the
number of halos from the catalogs again at 250 kpc.  We find that
dynamical friction noticeably affects the orbits of satellite halos,
in an apparent disagreement with the conclusion of
\citet{penarrubia_benson05}.  The KS test probability that our
calculated halo positions are drawn from the same distribution as the
$N$-body positions is 23\% with dynamical friction but only 2\%
without it.

Even though we do not reproduce the positions of the subhalos exactly,
we find that the calculated orbits are statistically consistent with
those in the simulation in the region of interest ($r < 200$~kpc).
This test therefore suggests that the motions of objects in a
hierarchically-forming galaxy can be reasonably accurately calculated
by approximating the total gravitational potential as the sum of
spherical halos.  We now apply this method to calculate the orbits of
globular clusters.

%Note that we do not use these halo orbits to calculate the orbits of
%model globular clusters.  Instead, as described in
%\S~\ref{sec:potential}, we use their positions in the $N$-body
%catalogs and interpolate between the output times.  The cluster
%orbits, which we discuss next and which we use to calculate the
%dynamical evolution, are therefore unaffected by this deviation of
%halo orbits from the $N$-body results.

Within the disks of progenitor galaxies, all clusters in our model begin
on nearly circular orbits.  Present globular clusters in the Galaxy
could either have formed in the main disk, have come from the
now-disrupted progenitor galaxies, or have remained attached to a
satellite galaxy.  Figure~\ref{fig:orbits_cl} shows the three
corresponding types of cluster orbits.  Even the clusters formed within
the inner 10 kpc of the main Galactic disk do not stay on circular
orbits.  They are scattered to eccentric orbits by accreted satellites,
while the growth of the disk increases the average orbital radius.
Triaxiality of the dark halo (not included in present calculations)
would also scatter the cluster orbits.  The clusters left over from the
disrupted progenitor galaxies typically lie at larger distances, between
20 and 60 kpc, and belong to the inner halo class.  Their orbits are
inclined with respect to the Galactic disk and are fairly isotropic.
The clusters still associated with the surviving satellite galaxies are
located in the outer halo, beyond 100 kpc from the Galactic center.
Note that these clusters may still be scattered away from their hosts
during close encounters with other satellites and consequently appear
isolated.

%%%%% Fig 2  %%%%%%

\begin{figure*}[!t]
\epsscale{1.2}  
\plotone{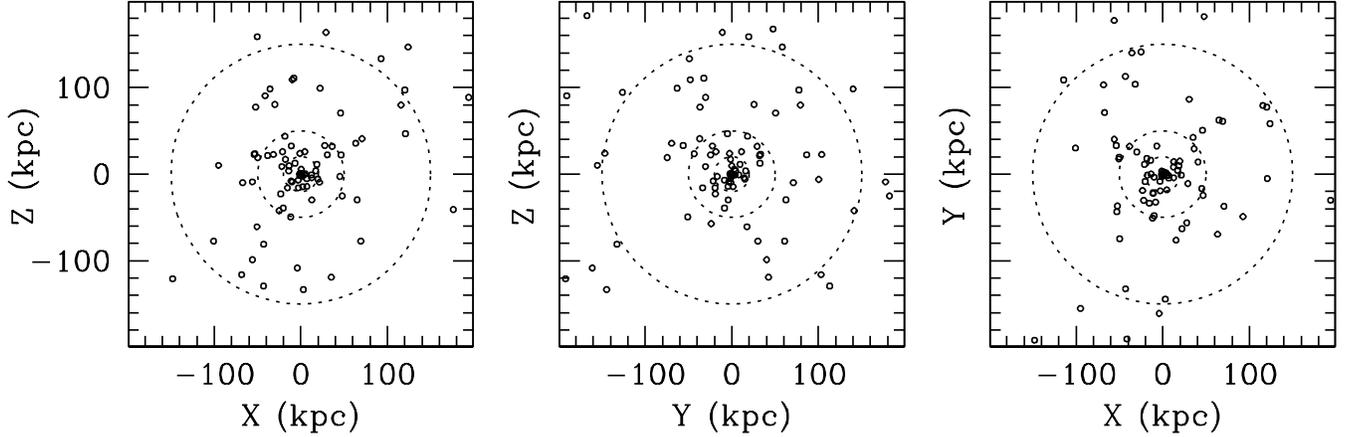}
\vspace{-0.5cm}
\caption{Spatial distribution of surviving clusters in the Galactic
  frame for model Sb-ii. Dotted circles are at projected
  distances of 20, 50 and 150~kpc.}
  \label{fig:positions}
\end{figure*}

%%%%% Fig 3  %%%%%%

\begin{figure*}[t]
\epsscale{1.0}  
\plotone{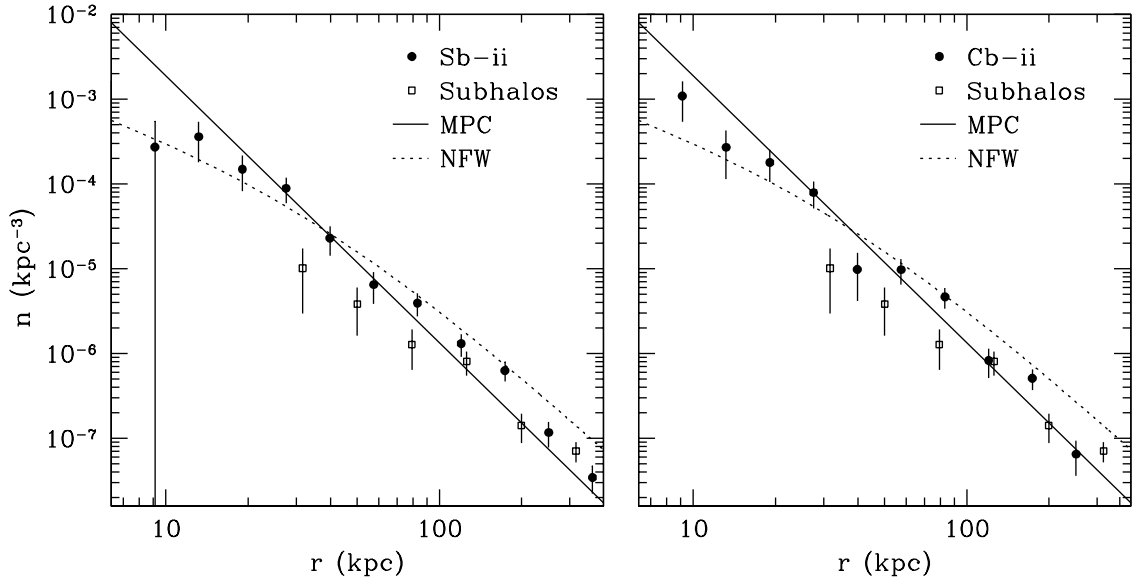}
\vspace{-0.5cm}
\caption{Number density profiles of surviving globular clusters ({\it
  filled circles}) in models Sb-ii ({\em left}) and Cb-ii ({\em
  right}) can be fit by the same power law, $n(r)\propto r^{-2.7}$.
  Vertical error bars are from Poisson statistics alone.  The
  distribution of model clusters is somewhat steeper than that of
  surviving satellite halos ({\it open circles}), but it is similar to
  the distribution of smooth dark matter ({\it dotted line}, arbitrary
  normalization).  It is also consistent with the observed
  distribution of metal-poor globular clusters in the Galaxy ({\it
  solid line}), plotted using the data from the catalog of
  \protect\citet{harris96}.}
  \label{fig:density}
\end{figure*}

%%%%%%% Fig 4 %%%%%%

\begin{figure*}[t]
\epsscale{0.8}
\plotone{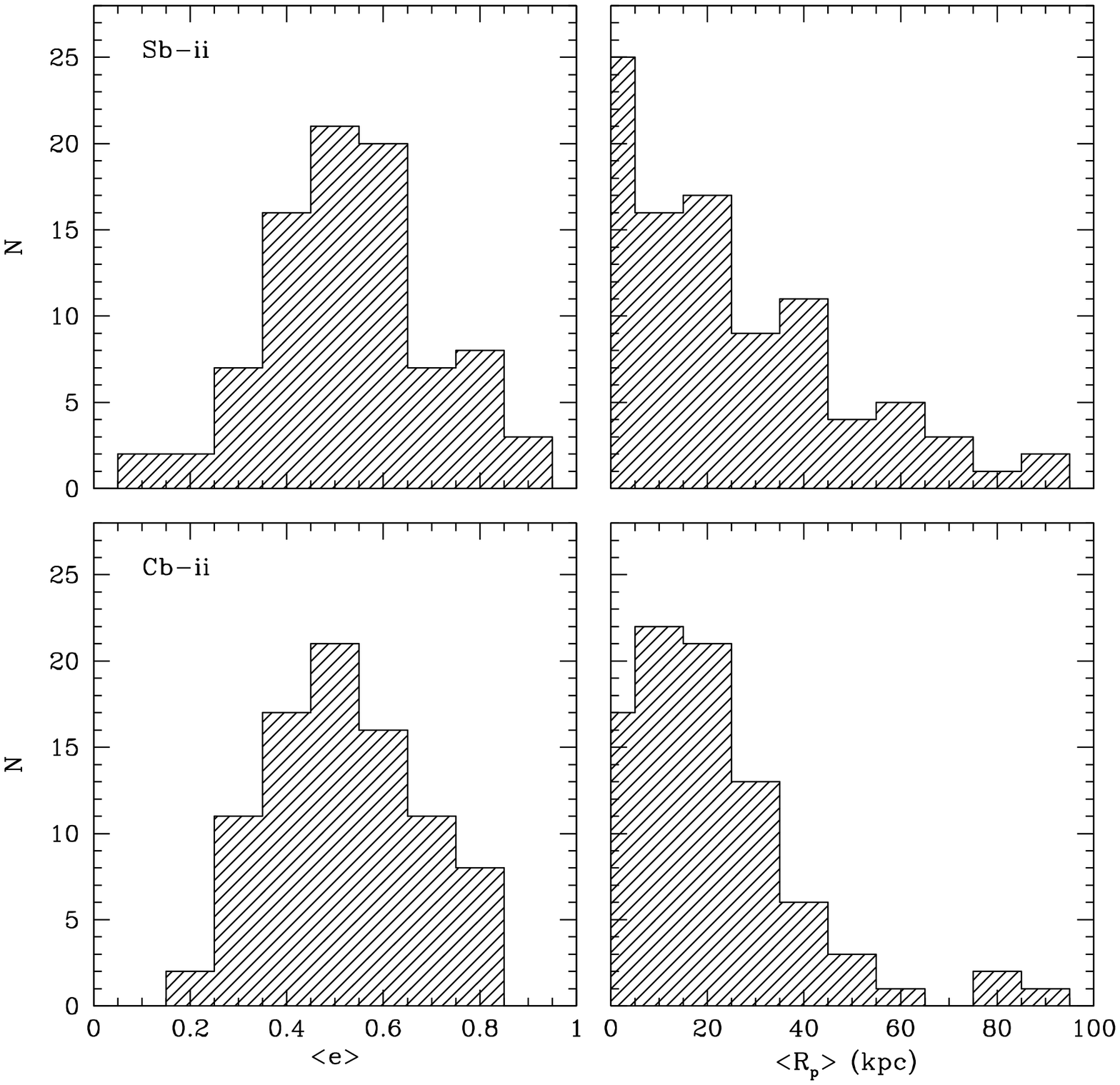}
\vspace{-0.3cm}
\caption{Mean eccentricity ({\em left}) and mean pericenter ({\em
  right}) distributions of the surviving globular clusters in models
  Sb-ii ({\em top}) and Cb-ii ({\em bottom}).
  \label{fig:ecc}}
\end{figure*}

%%%%%%%%%%%%%%%%%%%% 

\subsection{Spatial distribution}
  \label{sec:spatial}

Hierarchical mergers of progenitor galaxies ensure that the present
spatial distribution of the globular cluster system is roughly
spheroidal.  Figure~\ref{fig:positions} shows the positions of the
surviving globular clusters in the Galactic frame, for the best-fit
model Sb-ii described in \S~\ref{sec:mass_function}.  Most clusters
are now within 50 kpc from the center, but some are located as far as
200 kpc.  The median distance is 90 kpc, which is significantly larger
than the median distance of the metal-poor Galactic clusters (7 kpc).
Our simulated galaxy is perhaps twice as extended as the Milky Way,
but this alone is not enough to explain the discrepancy.  It could be
that some metal-poor clusters have formed later within the main disk,
which is not included in our current model.  

Figure~\ref{fig:density} shows that the azimuthally-averaged space
density of globular clusters is consistent with a power law,
$n(r)\propto r^{-\gamma}$.  The best models have logarithmic slopes
$\gamma = 2.65$ (Sb-ii) and $\gamma=2.72$ (Cb-ii).  The slopes for all
other models are given in Table~\ref{table:models1}.  Since all of the
distant clusters originate in progenitor galaxies and share similar
orbits with their hosts (disrupted and survived), the distribution of
the clusters is similar to the overall distribution of dark matter
given by the NFW profile.  The cluster profile is steeper than that of
the surviving satellite halos ($\gamma \approx 2.2$), because the
surviving satellites are preferentially found in the outer regions of
the main halo.  A higher fraction of the inner satellites is
disrupted, but the globular clusters that they hosted are dense enough
to survive the tidal field of the inner galaxy.

The density profile of model clusters is also similar to the observed
distribution of the metal-poor ($\mbox{[Fe/H]} < -0.8$) globular
clusters in the Galaxy, calculated using the online catalog of
\citet{harris96}: $\gamma=3.15 \pm 0.09$.  Such comparison is
appropriate, for our model of cluster formation at high redshift
currently includes only low metallicity clusters ($\mbox{[Fe/H]} \le
-1$).  Thus the formation of globular clusters in progenitor galaxies
with subsequent merging is consistent with the observed spatial
distribution of the Galactic metal-poor globulars.

We have also tested the effect of dynamical friction on the spatial
distribution of clusters in our best models.  We included this effect
on the orbits only within the main halo, but not in the original host
galaxies.  We again use Chandrasekahr's approximation for the drag
force (eq.~[\ref{eq:DF}]), with the appropriate coulomb logarithm,
$\ln{\Lambda_{\rm cl}}=12$, and replace the mass of the subhalos by
the initial mass of the clusters.  By using the initial cluster mass,
not reduced by subsequent dynamical evolution, we consider the maximum
effect that dynamical friction can have on cluster orbits.  Unlike the
case of the halo orbits, we find that dynamical friction does not
noticeably change the cluster distribution.  The density profile can
again be fit by a power law with the slope that differs from the case
without dynamical friction by only $\Delta\gamma = 0.02$.  This
difference is well within the errors of the power-law fit procedure
and is therefore not significant.  We note that the distributions of
subhalos with and without dynamical friction are also consistent
within the errors, with $\Delta\gamma = 0.06$.

%%%%%%% Fig 5 %%%%%%

\begin{figure*}[t]
\epsscale{1.0}
\plotone{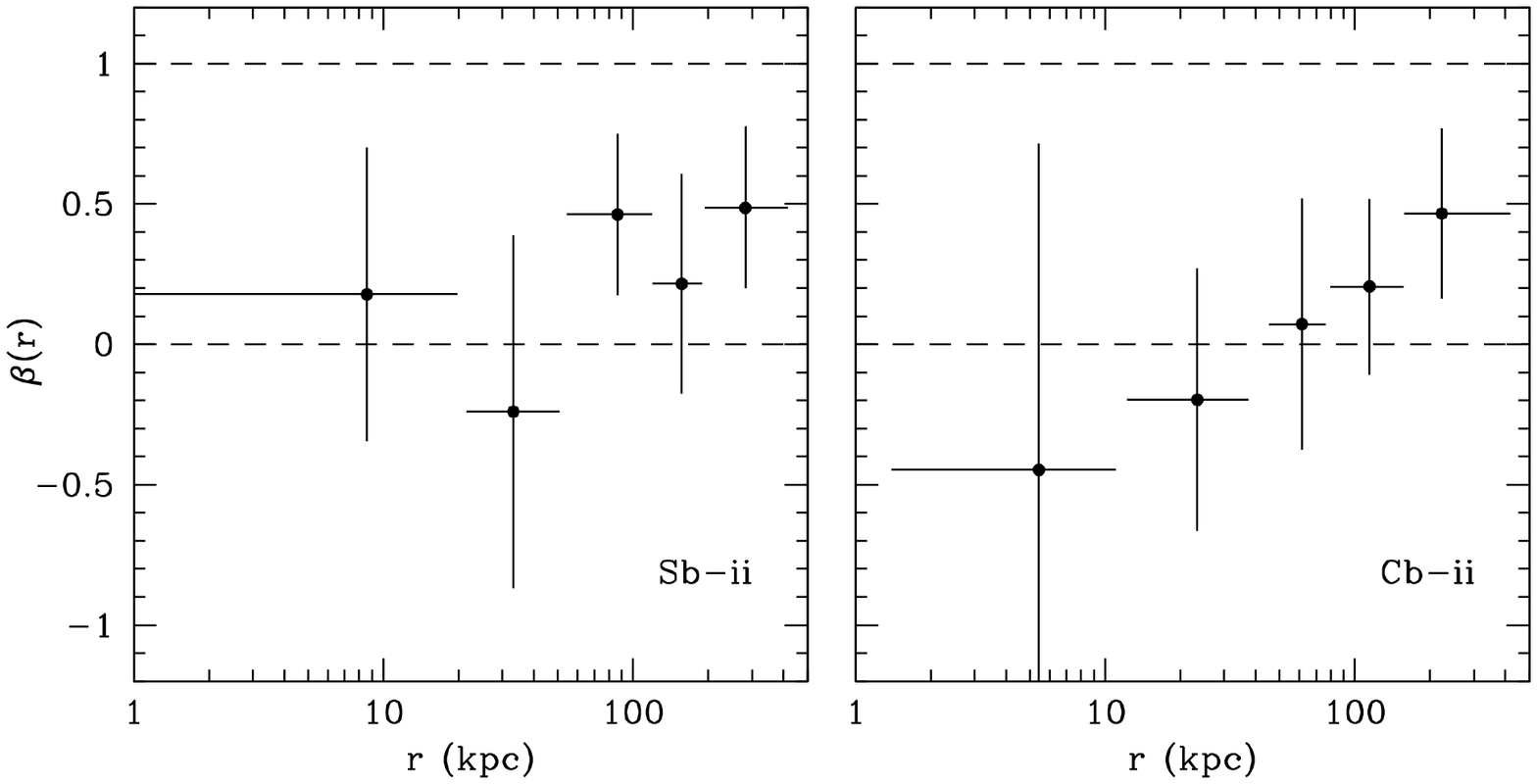}
\vspace{-0.5cm}
\caption{Anisotropy parameter $\beta$ as a function of radius for
  models Sb-ii ({\em left}) and Cb-ii ({\em right}).  The vertical
  error bars represent the error in the mean for each radial bin,
  while horizontal bars show the range of the bin.  Horizontal dashed
  lines illustrate an isotropic ($\beta=0$) and a purely radial
  ($\beta=1$) orbital distributions.}
  %The dotted line is the
  %anisotropy distribution adopted by \citet{fall_zhang01}, an
  %Eddington model with the anisotropy radius $R_A=5$~kpc.
  \label{fig:beta}
\end{figure*}

%%%%%%%%%%%%%%%%%%%%

Several studies in the literature looked at the {\it projected} spatial
distribution of globular clusters in nearby, mainly early-type galaxies.
\citet{kissler-patig97} find that the surface density profiles of
globular cluster systems in 15 elliptical galaxies are consistent with a
power law $\sigma(R) \propto R^{-\Gamma}$, where $1.0 < \Gamma < 2.2$.
More recent studies find similar results for elliptical galaxies:
$\Gamma = 1.2 - 1.9$ \citep{rhode_zepf04}, $\Gamma = 0.8 - 2.6$
\citep{puzia_etal04}, $\Gamma \approx 2.0$ \citep{harris_etal04},
$\Gamma \approx 1.4$ \citep{dirsch_etal05}, $\Gamma = 1.2 - 1.6$
\citep{forbes_etal06}, and for a spiral galaxy NGC~7814, which has
$\Gamma \approx 1.0$ \citep{rhode_zepf03}.  More luminous galaxies have
shallower globular cluster profiles, in agreement with an earlier study
by \citet{harris86}.  In the inner regions, many GC system profiles have
constant-density cores.  In addition, in all studied cases the
metal-poor globular cluster population is more spatially extended (has
lower $\Gamma$) than the metal-rich population.

% For a composite sample of globular clusters in nearby dwarf galaxies,
% \citet{sharina_etal05} find $\Gamma \approx 2.0$ for blue clusters in
% dSph galaxies and $\Gamma \approx 1.85$ for blue clusters in dIrr
% galaxies, in the inner projected 1 kpc from the dwarf galaxy center.
% These systems represent scaled-down versions of those in large spiral
% and elliptical galaxies.

To make a prediction for the surface density profile of our model
clusters, we project the cluster positions along random lines-of-sight
and fit a power-law to the resulting distribution.  After averaging over
6000 random projections, we find for the two best models: $\Gamma = 1.76
\pm 0.11$ (Sb-ii) and $\Gamma = 1.81 \pm 0.12$ (Cb-ii).  These power-law
slopes are consistent with the above-mentioned wide range of slopes of
the globular clusters systems in elliptical and some spiral galaxies.
These slopes $\Gamma$ are also consistent with the expectation for a
spherical distribution of model clusters, where $\Gamma = \gamma - 1$
and $\gamma = 2.7$.

\citet{forte_etal05} emphasize the similarity of the spatial
distribution of blue globular clusters in giant elliptical NGC 1399
with that inferred for dark matter, and \citet{cote_etal98} made a
similar case for M87.  As Figure~\ref{fig:density} shows, the density
profile of our model clusters is is similar to that of diffuse dark
matter, which is somewhat steeper than the density profile of the
surviving substructure \citep[e.g.,][]{nagai_kravtsov05,
maccio_etal06, weinberg_etal06}.  A similar connection has been made
by \citet{moore_etal06}, who identified globular cluster hosts with
high-density peaks in the dark matter distribution.  In our model, the
orbits of blue clusters are determined almost solely by the mass
assembly of the host galaxy and are therefore only weakly dependent on
the final morphological type of the galaxy.  Therefore, the similar
distribution of metal-poor clusters and dark matter is expected to
hold for both spiral and elliptical galaxies.  This is a robust
prediction of our model and is a direct consequence of globular
cluster formation in progenitor galaxies at high redshift.

In the continuous formation scenario, clusters that form early are
more centrally located than those that formed late.  We split the
clusters in model Cb-ii in three equal-size age bins and find that the
earliest 1/3 of the clusters have a median distance of 40 kpc, while
the latest 1/3 have a median distance of 77 kpc.  This result is also
a generic prediction of biased galaxy formation, where high density
peaks that collapse early have steeper and more concentrated density
profile than the more common peaks that collapse late
\citep[e.g.,][]{moore_etal06}.  The oldest metal-poor clusters should
therefore be found on the average closer to the center of the galaxy
than the somewhat younger ($\lesssim 2$ Gyr) metal-poor clusters.
Note that this does not apply to the metal-rich clusters, which
probably formed later in the main disk of the galaxy.

\subsection{Kinematics}
  \label{sec:kinematics}

From the positions and velocities of model clusters we can study their
kinematics at $z=0$.  We define orbital eccentricity using the
pericenter $R_{\rm p}$ and apocenter $R_{\rm a}$, the radial distances
from the center of the main halo at the closest and farthest points in
the orbit, respectively:
\beq 
  e = \frac{R_{\rm a} - R_{\rm p}}{R_{\rm a} + R_{\rm p}}.
  \label{eq:eccen}
\enq 
A typical cluster completes several revolutions around the center of the
main halo, with different $R_{\rm p}$ and $R_{\rm a}$ each time because
of the time-dependent nature of the potential.  We define the mean
eccentricity, $\left< e \right>$, by taking the average over all orbital
extrema between $z_{\rm acc}$ and $z=0$.

Figure~\ref{fig:ecc} shows the distribution of the average orbital
eccentricities and pericenter distances of surviving clusters in the two
best models.  Most orbits have moderate mean eccentricity, $0.4 < \left<
e \right> < 0.7$, expected for an isotropic distribution.  The
pericenters have a wide distribution, typically between 5 and 40 kpc
from the galactic center.  The eccentricities for our different models
averaged over all clusters are given in Table~\ref{table:models1}.  They
are consistent among different models and lie in the narrow range
between 0.52 and 0.58.

The anisotropy parameter is another important measure of the overall
kinematic properties:
\beq
  \beta(r) = 1 - \frac{\langle v_t^2 \rangle}{2 \langle v_r^2\rangle},
  \label{eq:beta}
\enq
where $\langle v_r^2 \rangle$ is the mean square radial velocity of
clusters in a radial bin, and $\langle v_t^2 \rangle$ is the mean square
tangential velocity.  By construction, $\beta=0$ corresponds to an
isotropic distribution, $\beta > 0$ is radially anisotropic, and $\beta
< 0$ is tangentially anisotropic.

Figure~\ref{fig:beta} shows the velocity anisotropy of the model
globular cluster system as a function of radius in the main halo at
$z=0$, for the two best models Sb-ii and Cb-ii.  In the inner 50 kpc
from the Galactic center, $\beta(r) \approx 0$, consistent with an
isotropic distribution.  At larger distances cluster orbits tend to be
more radial, reaching a maximum value $\beta_{\rm max} \approx 1/2$,
as is typical of dark matter satellites in $N$-body simulations
\citep[e.g.,][]{colin_etal00}.  There, in the outer halo, host
galaxies have had only a few passages through the Galaxy or even fall
in for the first time.  The anisotropy of cluster orbits in our model
derives directly from the orbits of the satellite galaxies.

%The dotted lines in Figure~\ref{fig:beta} show the anisotropy
%distribution assumed by \citet{fall_zhang01}, who used an Eddington
%model with the anisotropy radius $R_{A}=5$~kpc.  In their model, the
%orbits are isotropic at small radii and become completely radial at $r
%\gg R_{A}$.  

The velocity distribution of the model cluster system has a moderate
rotational component in the inner 50 kpc, $V_{\rm rot} \approx 56$ km
s$^{-1}$.  The distribution at larger radii is consistent with no
significant rotation.  The plane of rotation coincides with the plane
of the main disk.  Since we do not specify the sense of rotation of
the Galaxy when we construct a model for the gravitational potential,
we cannot determine whether the cluster motion is prograde or
retrograde with the respect to the disk.

The alignment of the rotation plane of the inner ``disk'' Galactic
clusters with the plane of the Galactic disk has always been expected
\citep{frenk_white80, thomas89} but could never be confirmed without
accurate measurements of the cluster proper motions.  Our models
provide the first indication that Galactic clusters may indeed rotate
along with the disk stars.  The estimated rotation velocity of the
metal-poor clusters is $30 \pm 25$ km s$^{-1}$ \citep{harris01}, which
is in reasonable agreement with $V_{\rm rot} = 56 \pm 25$ km s$^{-1}$
for the model clusters.

%%%%%%% Fig 6 %%%%%%%%

\begin{figure}[t]
\vspace{-0.3cm}
\centerline{\epsfysize3.5truein \epsffile{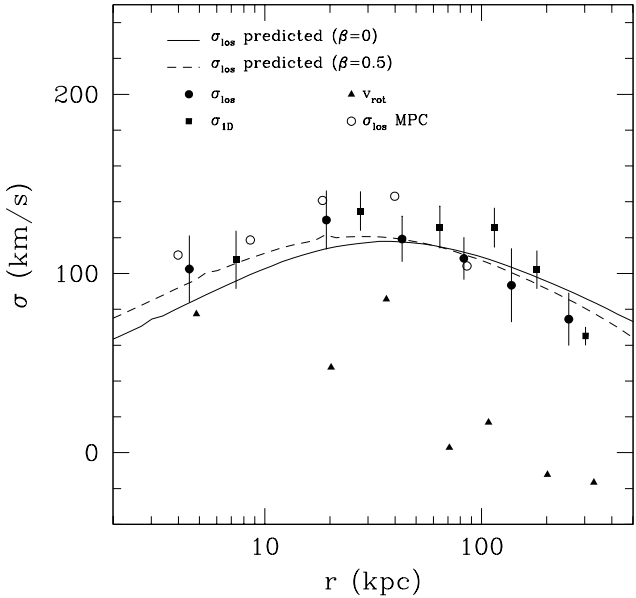}}
\vspace{-0.6cm}
\caption{Velocity dispersion profile of the surviving clusters in
  model Sb-ii.  We show the 1D ({\it filled squares}) and the
  projected line-of-sight ({\it filled circles}) velocity dispersions,
  as well as the rotational velocity of the cluster system ({\it
  filled triangles}) in equal-size radial bins.  Error bars are
  obtained using bootstrap resampling.  Lines show the predicted
  velocity dispersion in the analytical potential of the main dark
  matter halo, calculated using the spherical Jeans equation for a
  tracer population with the same radial distribution as model
  globular clusters, for isotropic orbits ($\beta=0$, {\it solid
  line}) and radially anisotropic orbits ($\beta=0.5$, {\it dashed
  line}).  Open circles show the line-of-sight velocity dispersion
  profile of Galactic metal-poor clusters.} 
  \label{fig:sigma}
\end{figure}

%%%%%%%%%%%%%%%%%%

Figure~\ref{fig:sigma} shows the rotation velocity and the velocity
dispersion profile in the model Sb-ii.  We calculate the line-of-sight
dispersion, $\sigma_{\rm los}$, by projecting the 3D velocities along
10000 random lines of sight and averaging the projected dispersion in
six equal-size radial bins.  Since the rotation velocity component is
small, subtracting it has a negligible effect on the velocity
dispersion, and therefore we do not include rotation when calculating
the velocity dispersion.  The projected dispersions, between 80 and
120 km s$^{-1}$, are almost identical to the spherical 1D velocity
dispersions, calculated as $\sigma_{1D} = \sigma_{3D}/\sqrt{3}$.  This
is yet another verification that the model clusters have an
approximately spherical and isotropic distribution.

Model clusters are also fair traces of the gravitational potential of
the dark matter halo.  Figure~\ref{fig:sigma} shows two projected
velocity dispersions for a tracer population with the radial density
profile of the model clusters, calculated from the Jeans equation
using a publicly available code {\sc Contra}%
\footnote{{\sc Contra} is available for download at\\ \mbox{\tt
http://www.astronomy.ohio-state.edu/$\sim$ognedin/contra/}.}
\citep{gnedin_etal04}.  The solid line is calculated assuming an
isotropic velocity distribution, while the dashed line is calculated
assuming a constant anisotropy parameter $\beta=0.5$.  If the model
clusters were in equilibrium with the spherical halo potential, in the
absence of rotation and of flattening of the disk potential, we would
expect the isotropic distribution to reproduce the model $\sigma_{\rm
los}$ at small radii and the anisotropic distribution to reproduce it
at large radii (see Figure~\ref{fig:beta}).  Both predicted curves
agree with the model $\sigma_{\rm los}$ within the errors, but the
anisotropic distribution appears to provide a better fit at all radii.
Figure~\ref{fig:sigma} also shows that the projected velocity
dispersion of model clusters agrees with that of metal-poor clusters
in the Galaxy, calculated using the \citet{harris96} catalog.

\section{Destruction of Globular Clusters}
  \label{sec:destruction}

The mass of a star cluster decreases with time as a result of several
dynamical processes of internal and external origin \citep{spitzer87}.
We consider the physical processes that have a major effect on the
cluster evolution and have been well studied in the literature: stellar
evolution, two-body relaxation, and tidal shocks.  As we note in
\S~\ref{sec:spatial} above, dynamical friction does not have a
noticeable impact on cluster orbits and therefore on their dynamical
evolution.

The differential equation that describes the cluster mass as a function
of time, $M(t)$, assuming that these effects are independent of each
other (a good first-order approximation), has the following form:
\beq 
  {dM \over dt} = -(\nu_{\rm se} + \nu_{\rm ev} + \nu_{\rm sh}) \, M,
  \label{eq:mass_evol}
\enq 
where $\nu_{\rm se}$, $\nu_{\rm ev}$, and $\nu_{\rm sh}$ are the
time-dependent fractional mass loss rates due to stellar evolution,
two-body relaxation, and tidal shocks, respectively
\citep[e.g.,][]{fall_zhang01}.  In the following subsections we describe
each term of eq.~(\ref{eq:mass_evol}) in detail.

%%%%%%% Fig 7 %%%%%%%

\begin{figure}[t]
\vspace{-0.3cm}
\centerline{\epsfysize3.5truein \epsffile{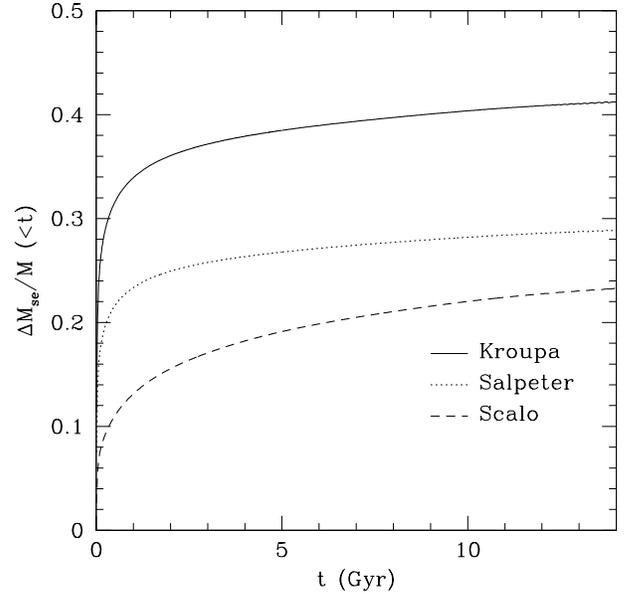}}
\vspace{-0.6cm}
\caption{Fractional mass loss from a globular cluster due to stellar
  evolution accumulated over time $t$, for three different IMFs:
  Salpeter ({\it dotted line}), Scalo ({\it dashed line}), and Kroupa
  ({\it solid line}; adopted in this work).}
  \label{fig:comp_mlse}
\end{figure}

%%%%%%%%%%%%%%%%%%%%

\subsection{Stellar Evolution}
  \label{sec:se}

We use a simple model to include the effect of mass loss due to stellar
evolution, following \citet{chernoff_weinberg90}.  All stars in the
cluster are assumed to form at the same time.  As stars of different
mass evolve away from the main sequence (MS), they lose mass through
stellar winds and supernovae explosions.  We assume that the lost
stellar material is not gravitationally bound to the cluster and escapes
immediately.

We adopt the stellar initial mass function (IMF) of \citet{kroupa01} 
between $0.1\, \Msun$ and $100\, \Msun$:
\beq
\frac{dN}{dm} \propto 
\left\{ \begin{array}{ll} 
  m^{-1.3}, & \quad  m < 0.5 \; \Msun \\ 
  m^{-2.3}, & \quad  m \ge 0.5 \; \Msun.
\end{array} \right. 
  \label{eq:stellar_IMF}
\enq
The mass loss in a given range of stellar masses ($m_1$, $m_2$) is: 
\beq
\Delta M(m_1,m_2)  = \int_{m_1}^{m_2} \frac{dN}{dm} \, (m-m_{f})\, dm,
  \label{eq:deltaM}
\enq
where $m_{f}$ is the remnant mass of a star with the initial mass $m$.
We adopt the relation between the initial mass and remnant mass given
by \citet{chernoff_weinberg90}.
%\beq
%    m_{f}   = 
%\left\{ \begin{array}{ll} 
%    0.58+0.22(m-1), & \quad m(\Msun) < 4.7 \\
%    0,              & \quad 4.7 < m(\Msun) < 8 \\
%    1.4,            & \quad 8 < m(\Msun) < 15 \\
%    0,              & \quad 15 < m(\Msun).
%  \end{array} \right.
%\label{eq:rel_m_mf} 
%\enq
%
We use a stellar evolution estimate of the MS lifetime for a star of
given initial mass \citep{chernoff_weinberg90, hurley_etal00} and
convert the mass loss in a range of masses to the mass loss in a range
of times, $\Delta M(m_1,m_2)=\Delta M(t_1,t_2)$.  Then we calculate the
fractional mass loss rate of a cluster as a function of time, $\nu_{\rm
se}(t)$, with sampling $\Delta t = 10^7$~yr.  We have tabulated and
spline-interpolated these values for use in the dynamical calculations.

Figure~\ref{fig:comp_mlse} shows the fractional mass loss due to stellar
evolution accumulated over time, $\int_0^t \nu_{\rm se}(t) dt$, for a
\citet{kroupa01} IMF and also for a \citet{salpeter55} and a
\citet{scalo86} IMFs, which are often used in galaxy modeling.  For our
adopted \citet{kroupa01} IMF, a cluster loses $\approx 30\%$ of its mass
in the first $3\times 10^8$ yr because of the evolution of massive stars
($m > 2 \, \Msun$) and an additional $\approx 10\%$ in the next 10~Gyr
due to the low mass stars leaving the MS.

Since the mass loss due to stellar evolution reduces the masses of all
clusters by the same fraction, this process does not affect the shape
of the GCMF and only shifts it towards lower masses.

\subsection{Two-body Relaxation}
  \label{sec:ev}

Internal two-body relaxation, a cumulative long-term effect of
gravitational interactions between stars that leads them to a Maxwellian
distribution of velocities, causes some stars to gain enough energy to
escape from the cluster.  The fractional mass loss due to two-body
relaxation can be approximated by the following expression
\citep{spitzer87}:
\beq
\nu_{\rm ev} = {\xi_e \over t_{\rm rh}} = 
  \frac{7.25 \xi_{e}\, \bar{m} G^{1/2}\, \ln{\Lambda_{\rm cl}}}
       {M^{1/2} R_h^{3/2}}
  \label{eq:nu_ev}
\enq 
where $\xi_e$ is the fraction of stars that escape per half-mass
relaxation time, $t_{\rm rh}$, $\bar{m}$ is the mean stellar mass, and
$\ln{\Lambda_{\rm cl}}$ is the usual Coulomb logarithm.

As a good approximation, we adopt constant values for various
quantities in equation~(\ref{eq:nu_ev}): $\xi_e = 0.033$, the mean
value obtained from $N$-body and Fokker-Planck models of cluster
evolution \citep{gnedin_etal99}, which is lower than the canonical
$\xi_e = 0.045$ obtained by \citet{henon61}; $\bar{m} \approx 0.87\,
\Msun$ for a Kroupa IMF (eq. [\ref{eq:stellar_IMF}]);
$\ln{\Lambda_{\rm cl}} = 12$, a typical value for globular clusters,
although it is expected to vary somewhat with time as the clusters
lose mass.  With these approximations, the time dependence of
$\nu_{\rm ev}$ is in the mass and half-mass radius of the cluster:
$\nu_{\rm ev}(t) \propto M(t)^{-1/2} R_h(t)^{-3/2}$.  Since we assume
that $R_h(t)$ depends only on $M(t)$ but does not depend explicitly on
cluster's position in the galaxy, the rate of two-body relaxation is
also independent of the location.  This has important consequences for
the lack of significant radial gradient of the cluster mass function.

\subsection{Tidal Shocks}
  \label{sec:sh}

Each time a cluster passes through the disk of its host galaxy or near
its nucleus, it experiences a rapid change of the external tidal force.
This change increases the average kinetic energy of stars, reducing the
cluster binding energy.  If the variation of the tidal field
(e.g. passage through the galactic disk) is fast compared with the
internal orbital period of stars in the cluster, the adiabatic
invariants of the orbit are not conserved causing some stars to become
unbound and escape from the cluster.  This is more effective in the
outer parts of the cluster where stars are weakly bound.

The effects of tidal shocks on the evolution of globular clusters have
been thoroughly studied in the literature \citep{ostriker_etal72,
spitzer87, kundic_ostriker95, gnedin_ostriker97, murali_weinberg97a,
murali_weinberg97b, murali_weinberg97c, gnedin_etal99b, gnedin_etal99}.
The physics is well understood and a general mathematical framework has
been developed to implement shocks due to the flattened (disk-like) and
spheroidal (bulge-like) components of the galactic potential in $N$-body
and Fokker-Planck simulations.  These models include adiabatic
corrections that ensure the conservation of orbital invariants when the
duration of the shock is long compared with the period of stars in the
cluster.  In such a case tidal shocking does not enhance the mass loss.

We implement tidal shocks using the semianalytical theory developed by
\citet{gnedin_etal99b} and \citet{gnedin03}.  The time variation of the
tidal force around a cluster can be divided into peaks (by absolute
value) surrounded by minima, and each peak is considered a separate
tidal shock.  The total tidal heating is the sum over all peaks.  The
tidal tensor is a symmetric matrix of the second spatial derivatives of
the time-dependent potential (eq.~[\ref{eq:total_potential}]), including
both the halo and disk components:
\beq
F_{\alpha\beta} = - \frac{\partial^2 \Phi}{\partial r_\alpha\, \partial r_\beta}.
  \label{eq:tidal_tensor}
\enq
A constant level of $F_{\alpha\beta}$ that does not vary with time is
not included in the integration; it is considered a steady tidal field
responsible for tidal truncation of clusters.  The ensemble-average
change in energy per unit mass at the half-mass radius can be
expressed as
\beq
\langle \Delta E_h \rangle
  = \frac{1}{2} \, \langle \Delta v \rangle^2
  = \frac{1}{6} \, I_{\rm tid} \, R_h^2,
  \label{eq:energy_change}
\enq   
where $I_{\rm tid}$ is the tidal heating parameter:
\beq
I_{\rm tid} \equiv \sum_{\alpha,\beta}
   \left(\int F_{\alpha \beta} \, dt\right)^2
   \left( 1 + \frac{\tau^2}{t_{\rm dyn}^2} \right)^{-3/2},
  \label{eq:tidal_heating}
\enq
where the sum extends over all components of the tidal tensor $\alpha,
\, \beta=\{x,y,z\}$, in Cartesian coordinates.  Here $\tau$ is the
duration of each tidal shock for each pair of coordinates $\alpha,
\,\beta$, and $t_{\rm dyn}$ is the dynamical time at the half-mass
radius.  We adopt a power-law adiabatic correction, appropriate for
relatively extended tidal shocks (see \citealt{gnedin_ostriker99} for
more detail).

A typical timescale for tidal heating to change the energy of stars at
the half-mass radius by the order of itself is
\beq
  t_{\rm sh} \equiv \frac{\vert E_h\vert }{(dE_h/dt)_{\rm sh}} 
    \simeq \frac{\vert E_h \vert }{2\langle \Delta E_h \rangle} P,
  \label{eq:shock_time}
\enq
where $P$ is the time interval between tidal shocks, which is
typically the orbital period of the cluster in its host galaxy.  The
factor of 2 in equation~(\ref{eq:shock_time}) comes from the
contribution of the second-order energy dispersion, $\langle\Delta
E^2\rangle$, which has been shown to be as important as the
first-order term, $\langle\Delta E\rangle$, for heating the cluster's
stars \citep{kundic_ostriker95}.  The energy per unit mass at the
half-mass radius can be approximated as $|E_h| \approx v_h^2/2 \approx
0.2\, GM/R_h$ \citep[e.g.,][]{spitzer87}.

Assuming that a fractional energy change in a tidal shock results in
the equal fractional mass loss \citep{chernoff_etal86}, we combine
equations (\ref{eq:energy_change}) and (\ref{eq:shock_time}) to derive
the mass loss rate due to tidal heating:
\beq
  \nu_{\rm sh} \simeq \frac{1}{t_{\rm sh}} 
                = \frac{5/3}{P}\, \frac{I_{\rm tid} R_h^3}{G M}.
  \label{eq:nu_sh}
\enq
The time dependence of this rate is different from that of two-body
relaxation: $\nu_{\rm sh}(t) \propto M(t)^{-1} R_h(t)^3$.  This
difference plays a critical role in shaping the mass function of
globular clusters.

%%%%%% Fig 8  %%%%%%%

\begin{figure}[t]
\vspace{-0.3cm}
\centerline{\epsfysize4.2truein \epsffile{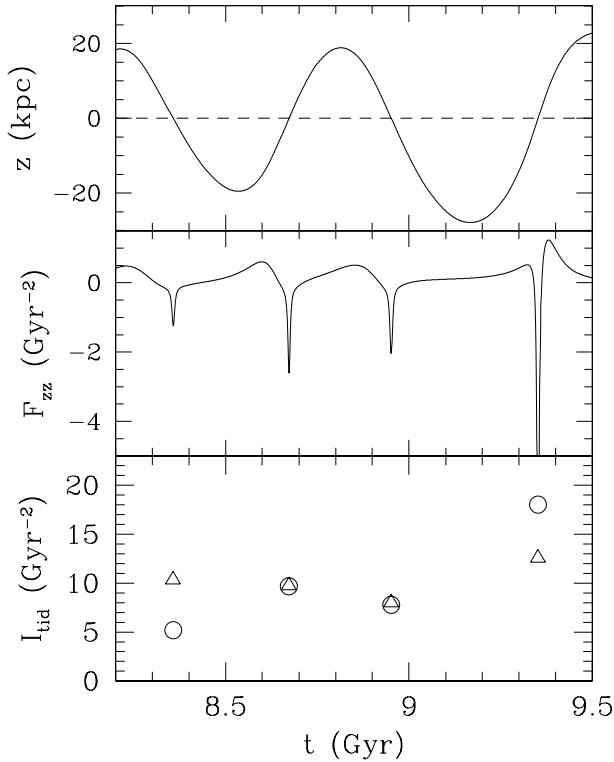}}
\vspace{-0.2cm}
\caption{An example of the calculation of disk shocking.  {\it Top
  panel:} vertical coordinate of the cluster with respect to the disk
  of the main galaxy.  {\it Middle panel:} main component of the tidal
  tensor, $F_{zz}$, responsible for disk shocking.  {\it Bottom
  panel:} comparison of the calculation of the tidal heating
  parameter, $I_{\rm tid}$, using the full tidal tensor (via
  eq. [\ref{eq:tidal_heating}], {\it circles}) and using a traditional
  parametrization of disk shocking (via eq. [\ref{eq:itid_disk}], {\it
  triangles}).}
  \label{fig:shocks}
\end{figure}

%%%%%%%%%%%%%%%%%%%%

Figure~\ref{fig:shocks} shows an example of our calculation of the
tidal heating parameter, $I_{\rm tid}$.  The cluster chosen for this
plot was formed in a satellite galaxy but was accreted by the main
halo at $t \approx 7$ Gyr.  It is now orbiting the main galaxy and
crosses the disk every 0.33 Gyr.  The dominant component of the tidal
tensor is $F_{zz}$, which describes the variation of the vertical
component of the tidal force during disk crossing.  The bottom panel
of Figure~\ref{fig:shocks} shows the comparison of the tidal heating
calculated via integration of equation (\ref{eq:tidal_heating}) with a
traditional way of parametrizing disk shocks.  Following
\citet{ostriker_etal72} and \citet{spitzer87}, we can write the
average energy change in one disk crossing as $\langle \Delta E_h
\rangle = 2 g_m^2 R_h^2 / 3 V_z^2$, where $g_m$ is the maximum
vertical gravitational acceleration dues ot the disk, and $V_z$ is the
component of cluster velocity perpendicular to the disk.  Comparing
this traditional expression with equation (\ref{eq:energy_change}) we
find that for disk shocking
\begin{equation}
  I_{\rm tid} \sim {4 g_m^2 \over V_z^2}.
  \label{eq:itid_disk}
\end{equation}
As Figure~\ref{fig:shocks} shows, for strong shocks the agreement of
the two methods is usually very good, whereas for a weak shock the
traditional expression overestimates the energy change by up to a
factor of 2.  Note that the adiabatic corrections, which we omitted
here in both cases for clarity, can significantly modify the actual
amount of heating and should always be included in the calculations.

%%%%%%%% Fig 9 %%%%%%%%%

\begin{figure*}[t]
\epsscale{0.9}  
\plotone{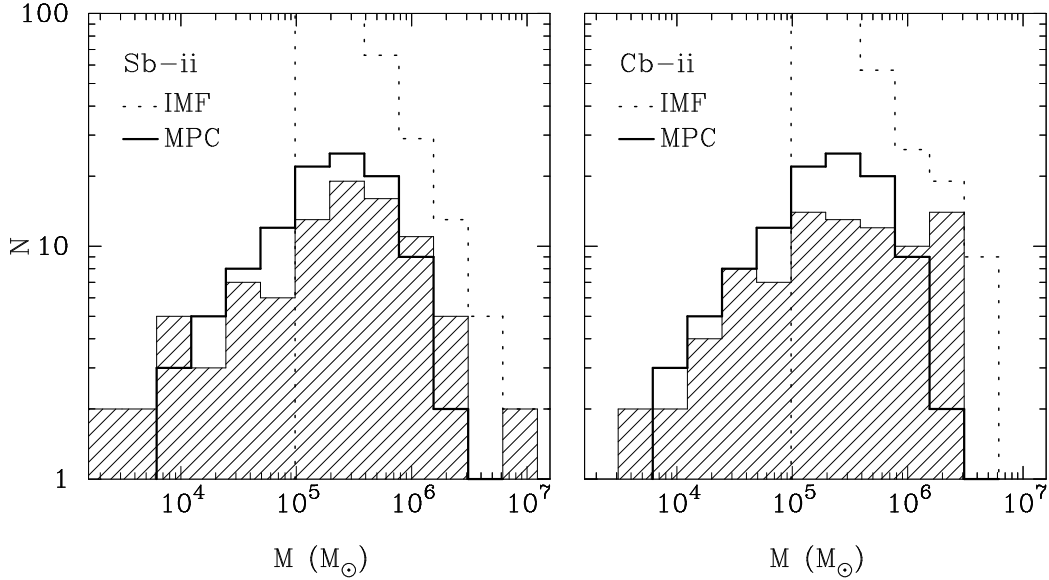}
\caption{Evolution of the mass function of model clusters from an
  initial power law ({\it dotted histogram}) to a peaked distribution
  at present ({\it shaded histogram}), including mass loss due to
  stellar evolution, two-body relaxation, and tidal shocks.  {\em Left
  panel:} model Sb-ii, {\em right panel:} model Cb-ii.  For
  comparison, the solid histogram shows the mass function of
  metal-poor globular clusters in the Galaxy.}
  \label{fig:mf}
\end{figure*}

%%%%%%%%%%%%%%%%%%%%%%%%

\section{Evolution of the Globular Cluster Mass Function}
  \label{sec:mass_function}

Using realistic cluster orbits and the specified $R_h - M$ relations, we
now integrate equation (\ref{eq:mass_evol}) for the evolution of GCMF
from $z=z_f$ to $z=0$.  Figure~\ref{fig:mf} shows the transformation
of the cluster mass function from an initial power law, $dN/dM \propto
M^{-2}$, into a final peaked distribution for models Sb-ii and Cb-ii.
High-mass clusters with $M \gtrsim 2\times 10^{5}\, \Msun$
approximately preserve the power-law shape of the IMF, while low-mass
clusters are more easily affected by the mass loss due to stellar
evolution, two-body relaxation, and tidal shocks.

We compare these model mass functions with the distribution of
metal-poor (MPC, ${\rm [Fe/H]} < -0.8$) globular clusters in the
Galaxy because they are the type of clusters described by our model
(see \S~\ref{sec:spatial}).  We obtain $V$--band luminosities of the
Galactic clusters from the current version of Harris's catalog
\citep{harris96} and use a constant mass-to-light ratio, $M/L_V=3$ in
solar units, to transform luminosities to stellar masses.  The two
plotted model mass functions are in excellent agreement with the
observed GCMF of the Galactic metal-poor clusters.

We fit the model GCMF with a traditional lognormal function, \\ $dN/dM
\propto \exp{[-(\log{M}-\log{M_{\rm peak}})^2/2\sigma^2]} \, dM$.  The
peak mass and the dispersion are very similar for both plotted models:
$M_{\rm peak} = (2.9\pm 0.13)\times 10^5\, \Msun$, $\sigma=0.61\pm
0.10$ (model Sb-ii) and $M_{\rm peak} = (2.3\pm 0.13)\times 10^5\,
\Msun$, $\sigma = 0.66 \pm 0.10$ (model Cb-ii).  For the metal-poor Galactic
clusters we find $M_{\rm peak} = 2.1\times 10^5\,\Msun$ and $\sigma =
0.51$.  The observed and model distributions are statistically
consistent with each other.

%%%%%%%% Fig 10  %%%%%%%%

\begin{figure*}[t]
\epsscale{0.9}  
\plotone{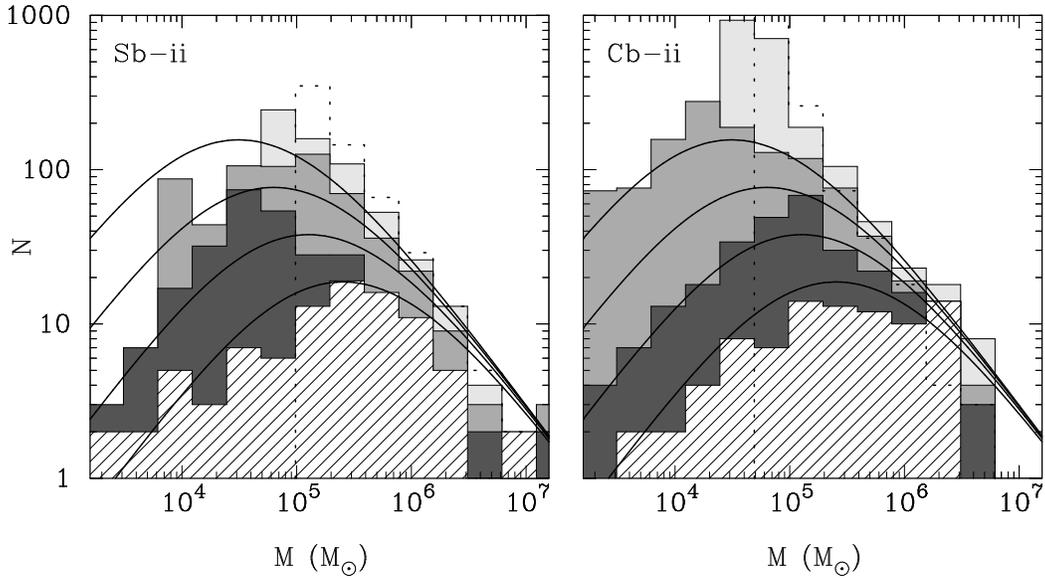}
\caption{Evolution of the mass function with time, beginning at $z=4$
  ({\it dotted histogram}) and subsequently after 1.5~Gyr ({\it light
  grey}), 3~Gyr, 6~Gyr ({\it dark grey}), and 12~Gyr ({\it shaded
  histogram}, corresponds to the present epoch).  {\em Left panel:}
  model Sb-ii, {\em right panel:} model Cb-ii.  For comparison, solid
  lines (same in both panels) show the predicted evolution of the mass
  function in the model of \citet{fall_zhang01} at 1.5~Gyr, 3~Gyr,
  6~Gyr, and 12~Gyr after cluster formation, from top to bottom.}
  \label{fig:mf_time}
\end{figure*}

%%%%%%%%%%%%%%%%%%%%

To further quantify this agreement, we use a Kolmogorov-Smirnov (KS)
test to calculate the probability of drawing the model GCMF and the
observed GCMF from the same underlying distribution.
Table~\ref{table:models1} lists the results for all models we consider.
The two models that we plot in Figure~\ref{fig:mf} and use to study the
spatial and kinematic distributions have the highest KS probabilities:
Sb-ii ($P_{KS}=0.24$) and Cb-ii ($P_{KS}=0.06$).  In these models the
initial half-mass density is the same for all massive clusters and it
remains constant through time: $R_h(t) \propto M(t)^{1/3}$.  A single
epoch of formation is slightly preferred to the continuous formation
scenario, but this difference is not statistically significant.  In
contrast, most of the other models are strongly inconsistent with the
observed mass function of Galactic metal-poor clusters.

Recall that in setting the initial distribution of model clusters, we
included only massive clusters with $M \ge 10^5\, \Msun$, as we expected
all lower-mass clusters to be disrupted.  We have verified this
assumption and find that the lowest initial mass of the clusters that
survive to $z=0$ are $3.8\times 10^5\, \Msun$ in model Sb-ii and $4.4
\times 10^5\, \Msun$ in model Cb-ii.  All of the low-mass clusters are
destroyed by the present time.

Table~\ref{table:models1} shows also the fractions of surviving model
clusters by mass and by number, for clusters with $M > 10^5\, \Msun$.
In a single formation epoch model Sb-ii, only $f_N = 0.16$ of the
initial massive clusters survive until $z=0$, but they contain a
higher fraction of the initial cluster mass, $f_M = 0.46$.  In the
continuous formation model Cb-ii, both fractions are a factor of $3-4$
lower: $f_N = 0.04$ and $f_M = 0.16$.  In the latter model, low-mass
clusters are relatively more abundant at $z_f > 4$, when the host
galaxies are smaller than at $z_f = 4$, and their subsequent rapid
disruption results in the smaller surviving fractions.

%%%%%%% Fig 11  %%%%%%%

\begin{figure}[t]
\plotone{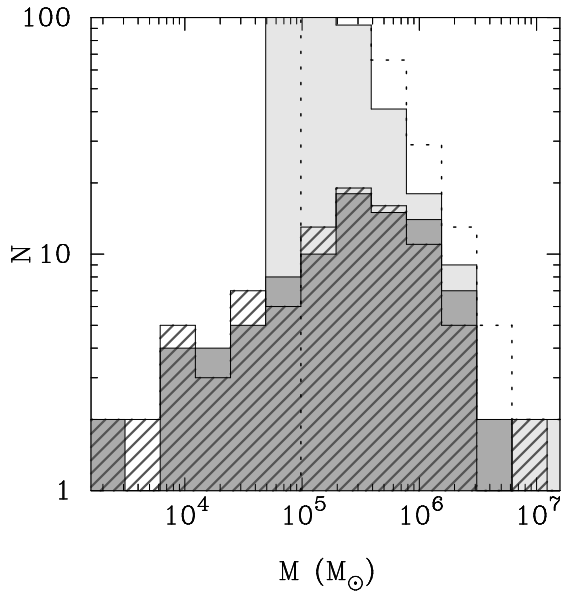}
\caption{Evolution of the mass function in model Sb-ii as a result of
  inclusion of different disruption processes: stellar evolution ({\it
  light-grey histogram}), two-body relaxation and evaporation ({\it
  dark histogram}), and tidal shocks ({\it hatched histogram}, final
  distribution).}
  \label{fig:mf_process}
\end{figure}

%%%%%%%%%%%%%%%%%%%%%%%

Figure~\ref{fig:mf_time} shows the evolution of the cluster mass
function with time, for the best models Sb-ii and Cb-ii.  In model
Sb-ii all clusters form at redshift $z=4$, but in model Cb-ii clusters
form continuously beginning at $z=9$.  To illustrate both formation
scenarios in the same way, we plot the distribution of clusters in
model Cb-ii at redshift $z=4$ (formed between $z=9$ and $z=4$) as the
``initial'' GCMF.  This epoch corresponds to a cosmic time of 1.5 Gyr,
which is almost exactly 12 Gyr ago, in our adopted cosmology.  The
sequence of histograms displays the progression of the GCMF at four
successive epochs (0, 1.5, 6, and 12~Gyr) after $z=4$.

For comparison, Figure~\ref{fig:mf_time} shows also the mass functions
predicted by the model of \citet{fall_zhang01} at the same epochs.
\citet{fall_zhang01} consider several initial forms of the GCMF,
several kinematic distributions of the clusters in the Galaxy, and
several prescriptions for disk shocking, and solve equation
(\ref{eq:mass_evol}) under these different assumptions.  They also
keep the half-mass density $M/R_h^3$ constant as a function of time
and find that this is the best way to reproduce the observed GCMF, in
agreement with our conclusions.  We select one of their models (top
left panel of their Fig. 3), which is the closest match to our models.
This model has an initial power-law GCMF with the slope $\alpha=2$ and
a single epoch of cluster formation, which we assume equal to $z=4$.
We normalize their model to have the same initial number of clusters
with $M > 10^5\, \Msun$ as in our model in each of the panels of
Figure~\ref{fig:mf_time}.

We find a reasonable agreement of our GCMF with the prediction of
\citet{fall_zhang01} at late times.  At 1.5 Gyr after cluster
formation, our clusters are less affected by the dynamical evolution,
especially in the continuous formation model Cb-ii.  Note that
although our models do not include clusters with $M < 10^5\, \Msun$
initially, this does not affect the comparison in the range $M = 10^5
- 4\times 10^5\, \Msun$, where our models predict somewhat slower
evolution than the model of \citet{fall_zhang01}.  The final
distributions in our two best models are similar to each other and to
the prediction of \citet{fall_zhang01}.  The peak of the log-normal
fit is $\log{M_{\rm peak}} = 5.46$ (Sb-ii) and 5.37 (Cb-ii) vs. 5.42
(FZ), and the dispersion is 0.61 (Sb-ii) and 0.66 (Cb-ii) vs. 0.63
(FZ).  Note that although we use the parameters of the log-normal
function for the ease of comparison between different models and
between models and observations, the mass function below the peak is
better described by a power law, $dN/d\log{M} \propto M$ or $dN/dM =
\mbox{const}$ \citep[see][]{fall_zhang01}.

Figure~\ref{fig:mf_process} provides insight into the effects of each
of the disruption mechanisms separately.  Mass loss due to stellar
evolution simply shifts the whole mass function to the left.  The
relative importance of two-body relaxation and tidal shocks depends on
the $R_h(M)$ relation.  Let us write this relation in a general
power-law form, $R_h \propto M^\delta$.  Then the ratio of the mass
loss rates is $\nu_{\rm ev}/\nu_{\rm sh} \propto M^{1/2} R_h^{-9/2}
\propto M^{1/2 - 9\delta/2}$.  Unless $\delta < 1/9$ (in the plotted
model Sb-ii, $\delta=1/3$), two-body relaxation is more effective in
low-mass clusters, and tidal shocks are more effective in high-mass
clusters.  The evolution of the $R_h(M)$ relation with time may also
affect the balance between the two processes.  The dark histogram in
Figure~\ref{fig:mf_process} shows how the inclusion of the evaporation
due to two-body relaxation removes a large number of low-mass clusters
and turns the mass function into a peaked distribution.  Inclusion of
tidal shocks affects mainly the most massive clusters and reduces
their surviving fraction.

%%%%%%%%  Fig 12  %%%%%%%%%

\begin{figure*}[t]
\epsscale{1.2}
\plotone{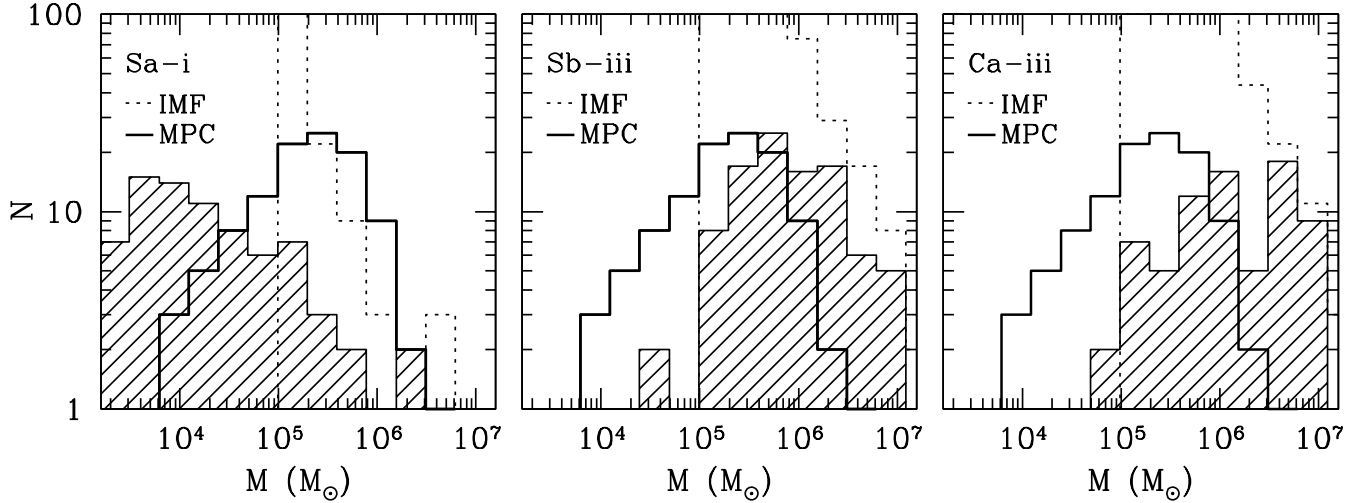}
\vspace{-0.8cm}
\caption{Models that fail to reproduce the observed mass function of
  metal-poor globular clusters: with $R_h(t) = {\rm const}$ (Sa-i, {\it
  left}), and with $R_h(t) \propto M(t)$ with single epoch of
  formation (Sb-iii, {\it middle}) and continuous formation (Ca-iii,
  {\it right}).  Normalization of the initial model distribution
  is such that the number of clusters at $z=0$ is close to the observed
  number of clusters in the Galaxy.}
  \label{fig:mf_fail}
\end{figure*}

%%%%%%%%%%%%%%%%%%%%%%%%%%

Our results agree with previous studies of the evolution of the
cluster mass function, which found that almost any initial function
can be turned into a peaked distribution as a consequence of the
dynamical evolution.  {\it The new result of our study is that not all
initial conditions and not all evolutionary scenarios are consistent
with the observed mass function.}

Figure~\ref{fig:mf_fail} provides three examples.  In the first (top
left panel, model Sa-i), the half-mass radius is kept fixed at the
median value for Galactic globulars, $R_h = 2.4$ pc, for clusters of all
masses and at all times.  The median density $M(t)/R_h^3$ decreases as
the clusters lose mass.  Therefore, two-body relaxation becomes less
efficient and spares many low-mass clusters, while tidal shocks become
more efficient and disrupt most high-mass clusters.  The final
distribution is severely skewed towards small clusters.  In the second
example (top right panel, model Sb-iii), the median density is initially
fixed, as in our best models, but the half-mass radius is assumed to
evolve in proportion to the mass, $R_h(t) \propto M(t)$.  In this case
the cluster density increases with time.  As a result, all of the
low-mass clusters are disrupted by the enhanced two-body relaxation,
while the high-mass clusters are unaffected by the weakened tidal
shocks.  The final distribution is skewed towards massive clusters.  In
the third example (bottom left panel, model Ca-iii), the initial
half-mass radius is the same for all clusters but then varies as $R_h(t)
\propto M(t)$.  In this model, which assumes a continuous formation
scenario from $z_f=9$ to $z_f=3$, the median cluster density again
increases with time and all of the low-mass clusters are destroyed by
two-body relaxation.  The initial relation $R_h(0)$ vs. $M(0)$ does not
appear to be as important for the final mass function as the evolution
of this relation with time, $R_h(t)$ vs. $M(t)$.

\section{Radial Variation of the Globular Cluster Mass Function}
  \label{sec:discussion}

A generic prediction of the dynamical disruption models is that the
evolution is faster in the inner parts of the galaxy, where the stronger
tidal field enhances both two-body relaxation and tidal shocks.  As a
result, the low-mass part of the mass function should be depleted more
strongly in the inner regions, causing an apparent peak mass to be
higher than in the outer regions \citep{ostriker_gnedin97}.  Such radial
dependence of the mass function is marginally present in the Galaxy
\citep{gnedin97} and more clearly in M31 \citep{barmby_etal01}, but it
has not been detected in other extragalactic systems.  This lack of the
observed radial variation has led \citet{fall_zhang01} to consider
anisotropic cluster orbits, which may suppress radial dependence of the
disruption processes.

Our best model Sb-ii does not predict a significant radial variation of
the mass.  To investigate this, we have divided the model sample into
two subsamples, splitting them at a distance $r = 30$ kpc from the
center.  The inner sample has 30 clusters, the outer 70 clusters.  (Had
we split the model sample into two equal parts, the dividing distance
would be 90 kpc, too large to expect any differences in the dynamical
evolution.)  A lognormal fit to the distribution of the inner clusters
gives the peak mass and dispersion $\log{M_{\rm peak}} = 5.56 \pm 0.20$,
$\sigma = 0.44 \pm 0.17$, while for the outer clusters $\log{M_{\rm
peak}} = 5.39 \pm 0.17$, $\sigma =0.82 \pm 0.13$, respectively.  The KS
probability of the two subsamples being drawn from the same distribution
is extremely high, $P_{KS} = 0.99$.

The mass functions of the model subsamples are consistent, within the
1$\sigma$ uncertainties, with the mass functions of the metal-poor
Galactic clusters, split in two equal-size parts at 7 kpc:
$\log{M_{\rm peak}} = 5.35\pm 0.09$, $\sigma = 0.40 \pm 0.07$ (inner
clusters), and $\log{M_{\rm peak}} = 5.31 \pm 0.16$, $\sigma = 0.61
\pm 0.11$ (outer clusters).  Note however the caveat that the assumed
dividing radius for our model clusters (30 kpc) is significantly
larger than that for the Galactic clusters (7 kpc).  This is
partially, but not entirely, offset by the fact that our model
galaxy is twice the size of the Milky Way and the extent of its
globular cluster system is expected to scale accordingly.

The lack of a significant radial gradient in our best models is due to
the contribution of tidal shocks being sub-dominant compared to the
contribution of two-body relaxation.  Under our assumption that the
median density $M/R_h^3$ is constant and does not depend on the
position in the galaxy, the rate of cluster mass loss due to two-body
relaxation alone is independent of location.  The rate of tidal shocks
does depend on the position and on cluster orbits, but because their
contribution is relatively small, the resulting radial gradient is
small as well.  We have verified this hypothesis by looking at one of
the failed models (Sa-i), where tidal shocks are strongly enhanced
compared to two-body relaxation.  This model shows a clear and strong
radial dependence of the mass function.  Therefore, the lack of the
gradient in our best models may be more of a product of the initial
set-up rather than a model prediction, but it does show that our best
models successfully pass this observational test.

\section{Conclusions}
  \label{sec:conclusions}

We have verified the hypothesis that the properties of globular
clusters formed within high-redshift disk galaxies are consistent with
the observed properties of Galactic metal-poor clusters.  After
calculating the orbits of model clusters in the time-variable
gravitational potential of a Milky Way-sized galaxy, we find that at
present the orbits are isotropic in the inner 50 kpc and
preferentially radial at larger distances.  All clusters located
outside 10 kpc from the center formed in satellite galaxies, some of
which are now tidally disrupted.  The spatial distribution of model
clusters is spheroidal, with a spherical density profile that can be
fit by a power law with the slope $\gamma \approx 2.7$, consistent
with the observations of Galactic metal-poor clusters.

Two-body relaxation dominates over other disruption processes and
drives the evolution of the cluster mass function from an initial
power law to the observed peaked distribution.  The evolution of the
mass function in our best models is in close agreement with the models
of \citet{fall_zhang01} at late times and is somewhat slower at early
times.  We find no significant variation of the mass function with
radius, which is largely due to the rate of two-body relaxation being
independent of cluster location.  Knowledge of realistic orbits is
important for accurate calculations of the globular cluster evolution,
although we find that tidal shocks play only a moderate role in
cluster disruption.  Therefore, any inaccuracies in orbit calculations
are not expected to significantly affect the evolution of the mass
function.

Interestingly, we find that not all initial conditions and not all
evolution scenarios are consistent with the observed mass function.
If the half-mass radius remains constant throughout the evolution
instead of decreasing with mass, the rate of two-body relaxation is
suppressed and the final mass distribution is skewed towards too many
low-mass clusters.  On the other hand, if the half-mass radius
decreases too quickly, proportionally to the mass, then two-body
relaxation is too efficient and there are no low-mass clusters left.
The successful models (Sb-ii and Cb-ii) require the average cluster
density, $M/R_h^3$, to be constant initially for clusters of all mass,
as in the model of \citet{kravtsov_gnedin05}, and to remain constant
with time.

We have investigated two formation scenarios.  Synchronous formation
of all clusters at a single epoch ($z=4$) and continuous formation
over a span of 1.6 Gyr (between $z=9$ and $z=3$) both appear
consistent with the observed mass function, spatial and kinematic
distributions of the Galactic globular clusters.  However, clusters
that formed at later epochs have more extended spatial distribution
than the clusters that formed at early epochs.  This is expected for
objects that trace the distribution of high-density peaks in
hierarchical galaxy formation.  If additional clusters form after
$z=3$, they are expected to have a shallower density profile than the
clusters we study in this paper.  The profile of our clusters is
consistent with, but is already somewhat shallower than the profile of
the Galactic metal-poor clusters.  Therefore, we do not expect a
significant fraction of metal-poor clusters to have formed after
$z=3$.  Metal-rich clusters, associated with the Galactic disk, are
not included in our present formation model and will be studied in
future work.

In online Tables~\ref{table:catalog1} and \ref{table:catalog2} we
provide catalogs of the physical properties of model clusters that
survive to the present time, for the best models Sb-ii and Cb-ii,
respectively.  These catalogs can be used to compare with other models
of the dynamical evolution, to study selection effects in
extragalactic systems, and to model the kinematics of globular
cluster systems.

\acknowledgements 
We thank the Ohio Supercomputer Center for the use of a Cluster Ohio
Beowulf cluster in conducting this research.  We thank Mike Fall, Andy
Gould, Andrey Kravtsov, Jeremy Tinker, and David Weinberg for helpful
discussions and comments on the manuscript.  J.~L.~P. was supported in
part by an OSU Astronomy Department Fellowship. O.~Y.~G. was supported
by NASA ATP grant NNG04GK68G and NSF grant AST-0407125.

\bibliography{gc}

%%%%%%%%%%%%%%%%%%%%%%%%%%%%%%%%%%%%%%%
%              Tables
%%%%%%%%%%%%%%%%%%%%%%%%%%%%%%%%%%%%%%%

\begin{deluxetable}{lllccccccl}
\tablecaption{\label{table:models1}}
\tablehead{
\colhead{Model} & \colhead{$R_h(0)$} & \colhead{$R_h(t)$} & \colhead{$\gamma$\tablenotemark{a}} & \colhead{$\langle e\rangle$\tablenotemark{b}} & \colhead{$f_M$\tablenotemark{c}} & \colhead{$f_N$\tablenotemark{d}} & \colhead{$\log{M_{\rm peak}}$\tablenotemark{e}} & \colhead{$\sigma$\tablenotemark{f}} &\colhead{$P_{KS}$\tablenotemark{g}}  
} 
\startdata
Sa-i   & const        & const        & 2.6 & 0.53 & 0.29 & 0.54 & 3.76 & 1.02 & $< 10^{-10}$ \\
Sa-ii  & const        & $M(t)^{1/3}$ & 2.6 & 0.54 & 0.40 & 0.20 & 5.02 & 0.68 & $2\times 10^{-4}$\\
Sa-iii & const        & $M(t)$       & 2.6 & 0.58 & 0.38 & 0.05 & 5.98 & 0.66 & $7\times 10^{-8}$\\
Sb-i   & $M(0)^{1/3}$ & const        & 2.6 & 0.53 & 0.45 & 0.27 & 4.97 & 0.95 & 0.0035 \\
Sb-ii  & $M(0)^{1/3}$ & $M(t)^{1/3}$ & 2.7 & 0.53 & 0.46 & 0.16 & 5.46 & 0.61 & 0.24 \\
Sb-iii & $M(0)^{1/3}$ & $M(t)$       & 2.6 & 0.57 & 0.54 & 0.09 & 5.88 & 0.52 & $4\times 10^{-10}$\\
Ca-i   & const        & const        & 2.5 & 0.54 & 0.15 & 0.43 & 1.96 & 1.40 & $<10^{-10}$ \\
Ca-ii  & const        & $M(t)^{1/3}$ & 2.7 & 0.53 & 0.11 & 0.05 & 5.14 & 0.71 & 0.023 \\
Ca-iii & const        & $M(t)$       & 2.7 & 0.57 & 0.16 & 0.02 & 6.04 & 0.62 & $<10^{-10}$ \\
Cb-i   & $M(0)^{1/3}$ & const        & 2.8 & 0.52 & 0.17 & 0.08 & 4.88 & 0.83 & $3\times 10^{-4}$\\
Cb-ii  & $M(0)^{1/3}$ & $M(t)^{1/3}$ & 2.7 & 0.52 & 0.15 & 0.04 & 5.37 & 0.66 & 0.063 \\
Cb-iii & $M(0)^{1/3}$ & $M(t)$       & 2.7 & 0.56 & 0.21 & 0.03 & 6.08 & 0.51 & $<10^{-10}$ \\
\enddata
\tablenotetext{a}{Power law slope of the number density distribution:
  $n(r)\propto r^{-\gamma}$.}
\tablenotetext{b}{Time-averaged eccentricity of the surviving clusters.}
\tablenotetext{c,d}{Fraction of the mass and fraction of the number of
  surviving globular clusters with respect to the initial GCMF, for
  clusters with $M > 10^5\, \Msun$.  All clusters with the initial
  mass $M < 10^5\, \Msun$ are disrupted.}
\tablenotetext{e,f}{Peak mass and dispersion (in log scale) of a
  lognormal distribution fitted to the mass function of model clusters.}
\tablenotetext{g}{Kolmogorov-Smirnov probability of the mass function
  of model clusters being drawn from the same distribution as the
  Galactic metal-poor globular clusters.}

\end{deluxetable}

%%%%%%%%%%%%%%%%%%%%%%%%%%%%%%%%%%%%%%%

\begin{deluxetable}{ccccccccrrrrrr}
\tablewidth{0pt}
\tabletypesize{\small}
\tablecaption{Catalog of clusters in model Sb-ii. \label{table:catalog1}}
\tablehead{
\colhead{$z_{\rm f}$} & \colhead{$\log{M}$} & \colhead{$\log{M_{i}}$} & \colhead{$R_{h}$} & \colhead{$R_{h,i}$} & \colhead{$r$} & \colhead{$r_{i}$} & \colhead{$\log{M_{{\rm halo},i}}$} & \colhead{$x$} & \colhead{$y$} & \colhead{$z$}  & \colhead{$v_{x}$} & \colhead{$v_{y}$} & \colhead{$v_{z}$}
} 
\startdata
 4.0 & 5.72 & 6.10 &  2.5 &  3.4 &   4.2 &   0.7 & 10.9 & $-$3.8 & $-$1.6 & $-$0.6 & $-$25.4 & $-$12.3 & $-$28.5 \\ 
 4.0 & 5.34 & 5.91 &  1.9 &  2.9 &   6.4 &   1.1 & 10.9 & $-$0.4 & $-$2.9 & $-$5.7 & $-$28.5 & $-$144.7 & $-$21.7 \\ 
 4.0 & 3.99 & 5.72 &  0.7 &  2.5 &   1.4 &   1.2 & 10.9 & 1.1 & 0.5 & 0.7 & 57.5 & 4.0 & 18.6 \\ 
 4.0 & 5.15 & 5.87 &  1.6 &  2.8 &   1.4 &   1.5 & 10.9 & 0.8 & 0.1 & 1.1 & 23.1 & 112.9 & 26.8 \\ 
 4.0 & 6.18 & 6.43 &  3.6 &  4.3 &   0.4 &   1.8 & 10.9 & $-$0.4 & $-$0.1 & $-$0.0 & $-$110.2 & $-$3.7 & $-$94.9 \\ 
 &&&&&&& ...
\enddata
\tablecomments{Columns are: $z_{f}$ -- redshift of formation,
  $\log{M}$ -- present cluster mass in $\Msun$, $\log{M_{i}}$ --
  initial mass, $R_{h}$ -- present half-mass radius of the cluster in
  pc, $R_{h,i}$ -- initial half-mass radius, $r$ -- present distance
  to the center of the main galaxy in kpc, $r_i$ -- initial distance
  to the center of the main galaxy, $\log{M_{{\rm halo},i}}$ -- mass
  of the host galaxy at the time of cluster formation in $\Msun$, $x$,
  $y$, $z$ -- present coordinates of the cluster with respect to the
  center of the main galaxy in kpc, $v_{x}$, $v_{y}$, $v_{z}$ --
  present velocities of the cluster in km s$^{-1}$.  Complete table is
  available online.}
\end{deluxetable}

%%%%%%%%%%%%%%%%%%%%%%%%%%%%%%%%%%%%%%

\begin{deluxetable}{ccccccccrrrrrr}
\tablewidth{0pt}
\tabletypesize{\small}
\tablecaption{Catalog of clusters in model Cb-ii. \label{table:catalog2}}
\tablehead{
\colhead{$z_{\rm f}$} & \colhead{$\log{M}$} & \colhead{$\log{M_{i}}$} & \colhead{$R_{h}$} & \colhead{$R_{h,i}$} & \colhead{$r$} & \colhead{$r_{i}$} & \colhead{$\log{M_{{\rm halo},i}}$} & \colhead{$x$} & \colhead{$y$} & \colhead{$z$}  & \colhead{$v_{x}$} & \colhead{$v_{y}$} & \colhead{$v_{z}$}
}
\startdata
 6.2 & 4.91 & 5.83 &  1.3 &  2.7 &   2.3 &   1.4 & 10.5 & 0.7 & 2.1 & 0.7 & 35.3 & 32.5 & 37.7 \\ 
 5.6 & 4.96 & 5.83 &  1.4 &  2.7 &   3.6 &   2.8 & 10.5 & 0.6 & 3.5 & 1.2 & 123.9 & 25.4 & 64.7 \\ 
 5.4 & 5.01 & 5.83 &  1.5 &  2.7 &   2.6 &   0.7 & 10.6 & $-$1.3 & $-$1.9 & $-$0.7 & $-$94.4 & $-$30.2 & $-$34.0 \\ 
 5.2 & 5.75 & 6.13 &  2.6 &  3.4 &   9.8 &   1.7 & 10.7 & 4.2 & 7.7 & 2.7 & 42.5 & 7.9 & 84.2 \\ 
 5.0 & 5.68 & 6.08 &  2.4 &  3.3 &  17.3 &   6.5 & 10.7 & 3.5 & 16.5 & 4.8 & 79.0 & 54.0 & 10.1 \\
 &&&&&&& ...
\enddata
\tablecomments{Columns are the same as in Table~\ref{table:catalog1}.
  Complete table is available online.}
\end{deluxetable}

\end{document}